\documentclass[reprint, superscriptaddress, pra, aps, nolongbibliography, amsmath, amssymb, floatfix]{revtex4-2}

\usepackage{graphicx}
\usepackage{dcolumn}
\usepackage{bm}
\usepackage[normalem]{ulem}
\usepackage{braket}
\usepackage{float}
\usepackage{amsmath,amssymb,bm}
\usepackage{ascmac}
\usepackage{cases}	       
\usepackage{array}
\usepackage{ulem}
\usepackage{hyperref}
\hypersetup{%
colorlinks=True,
linkcolor=blue, 
citecolor=blue,
urlcolor=blue
}
\renewcommand{\sout}[1]{}
\graphicspath{{figure/}}

\begin{document}

\newcommand{\mrm}[1]{\mathrm{#1}}
\newcommand{\mcal}[1]{\mathcal{#1}}
\newcommand{\dif}[2]{\frac{\mathrm{d} #1}{\mathrm{d} #2}}
\newcommand{\abs}[1]{\left\lvert #1 \right\rvert}
\newcommand{\ketbra}[2]{\ket{#1}\!\bra{#2}}

\bibliographystyle{apsrev4-2}

\title{Non-demolition fluorescence readout and high-fidelity unconditional reset of a fluxonium qubit via dissipation engineering}

\author{Shu Watanabe}
 \email{watanabe@qipe.t.u-tokyo.ac.jp}
 \affiliation{%
 Department of Applied Physics, Graduate School of Engineering, The University of Tokyo, 7-3-1 Hongo, Bunkyo-ku, Tokyo 113-8656, Japan
}%

\author{Kotaro Hida}
 \affiliation{%
 Department of Applied Physics, Graduate School of Engineering, The University of Tokyo, 7-3-1 Hongo, Bunkyo-ku, Tokyo 113-8656, Japan
}%

\author{Kohei Matsuura}
 \email{matsuura@qipe.t.u-tokyo.ac.jp}
 \affiliation{%
 Department of Applied Physics, Graduate School of Engineering, The University of Tokyo, 7-3-1 Hongo, Bunkyo-ku, Tokyo 113-8656, Japan
}%

\author{Yasunobu Nakamura}%
\affiliation{%
 Department of Applied Physics, Graduate School of Engineering, The University of Tokyo, 7-3-1 Hongo, Bunkyo-ku, Tokyo 113-8656, Japan
}%
\affiliation{%
 RIKEN Center for Quantum Computing, Wako, Saitama 351-0198, Japan
}%

\begin{abstract}
Non-demolition readout and high-fidelity unconditional reset of qubits are key requirements for practical quantum computation. For superconducting qubits, readout and reset are typically realized using resonators based on dispersive approximation. However, violations of the approximation often lead to detrimental effects on the qubit. In this work, we demonstrate non-demolition fluorescence readout and high-fidelity unconditional reset of a fluxonium qubit via dissipation engineering without employing a resonator. We design and implement a planar filter to protect the qubit from energy relaxation while enhancing the relaxation of the readout transition. By appropriately selecting the readout transition, we achieve fluorescence readout with enhanced quantum non-demolitionness. We also realize fast, high-fidelity and all-microwave unconditional reset of the fluxonium qubit. These results highlight the potential of superconducting-quantum computing architectures without relying on dispersive interaction between qubits and resonators.
\end{abstract}

\maketitle

\section{Introduction} \label{sec:introduction}
Superconducting quantum circuits are among the most promising physical platforms for realizing quantum computers~\cite{Kjaergaard2020, Bravyi2022}. Various types of superconducting qubits have been proposed and explored, each offering different advantages in terms of coherence, scalability, and control~\cite{Nakamura1999, Mooij1999, Koch2007, Yan2016, Gyenis2021}. Although extensive research has primarily focused on transmon qubits with simple circuit structures~\cite{Arute2019, Jurcevic2021, Acharya2024}, fluxonium qubits have recently attracted significant attention as an alternative~\cite{Manucharyan2009, Nguyen2022}. Fluxonium qubits exhibit coherence times exceeding 1~ms at the optimal external magnetic flux bias, commonly referred to as the sweet spot~\cite{Somoroff2023, FeiWang2024}. Moreover, their large anharmonicity and diverse transition degrees of freedom offer potential advantages at the physical layer of quantum computing architectures. To overcome challenges in conventional quantum-operation schemes, novel approaches leveraging the unique characteristics of fluxoniums have been proposed~\cite{Rower2024, Lin2024, Stefanski2024, Schirk2024, Lin2025}.

Quantum non-demolition~(QND) readout of a qubit is a fundamental prerequisite for practical quantum computation. In superconducting quantum circuits, a resonator far-detuned from the qubit is commonly used to mediate the interaction between the qubit and the environment for the readout~\cite{Blais2021}. An additional resonator is also used as a filter to protect the qubit from energy relaxation through the readout resonator~\cite{Reed2010}. This qubit readout via a resonator, referred to as dispersive readout, is based on perturbation called dispersive approximation~\cite{Blais2004}. However, the approximation does not always hold when fast readouts are targeted. It has been shown that a phenomenon called ``ionization'' or measurement-induced state transition~(MIST) occurs in transmon qubits 
 when the photon number in the resonator reaches a certain threshold, causing the qubit to transition to higher excited states~\cite{Sank2016, Shillito2022, Khezri2023, Nesterov2024, Dumas2024}. State transitions associated with dispersive readout have also been reported in fluxoniums~\cite{Vool2014, Kou2018, Ficheux2021, Gusenkova2021, Somoroff2023, Ding2023}. Notably, it has been reported that the transition probability from the ground states to the excited states of fluxoniums can be larger than that of the reverse transition~\cite{Ficheux2021, Somoroff2023, Ding2023}.

An unconventional readout scheme without using a resonator has been proposed and demonstrated for a fluxonium~\cite{Cottet2021}. This scheme is based on two types of transitions: a `dark' transition with a sufficiently long lifetime, used as the computational basis, and a `bright' transition that emits fluorescence photons conditioned on the qubit state, used as the readout transition. We can readout the qubit state by driving the readout transition and measuring the reflected signal. This readout scheme using conditional fluorescence is analogous to electron-shelving readout commonly used in atomic physics~\cite{Nagourney1986,Blatt1988,Monroe1995,Schmidt2005,Hume2007}. In this fluorescence readout, a qubit is directly coupled to the environment, with no resonator mediating the interaction between them. The absence of resonator degrees of freedom significantly reduces the difficulty in modeling and analyzing the readout process and identifying the cause of non-QNDness. In the previous work~\cite{Cottet2021}, however, the readout transition did not form a closed cycle, and the QNDness of the readout was primarily limited by the undesired relaxation channel from the readout subspace to the non-computational subspace. In addition, the requirement that the external decay rate of the readout transition must be sufficiently larger than that of the computational subspace was satisfied using a three-dimensional copper waveguide with a specially designed input and output ports. The use of such three-dimensional structures limited the scalability and integrability of circuits. 

Another essential requirement for quantum computation is the ability to unconditionally reset qubits. Fast and high-fidelity unconditional reset not only mitigates the impact of state-assignment errors and non-QND errors of the measurement but also helps reduce uncorrectable leakage errors on fault-tolerant quantum computing. Active unconditional reset of fluxoniums is typically implemented using resonators dispersively coupled to the qubits to exchange the qubit excitation with the resonator excitation that is immediately emitted to the environment. Previous studies on active reset with resonators can be broadly categorized into two types. The first type uses a resonator-assisted Raman process, similar to that for transmons~\cite{Magnard2018, Sunada2022, Zhang2021}. Here, a qubit can be reset by simultaneously applying two microwave drives to form a $\Lambda$ system in the qubit--resonator system. 
While this scheme is all-microwave, strong driving is required for fast reset since the scheme involves a second-order process. The other type uses a sideband transition~\cite{Bao2022, Wang2024}. By applying a fast flux-bias pulse to offset the qubit from the sweet spot, one can drive the sideband transition between the single-excitation manifold in the qubit--resonator system, which is forbidden at the sweet spot. While a large flux offset enables fast and efficient reset, the use of fast flux-bias pulses causes additional decoherence and requires careful compensation for the pulse distortion~\cite{Rol2020,Sung2021,Li2024}. 

\begin{figure}[t]
    \centering
    \includegraphics[width=8.6cm, pagebox=cropbox, clip]{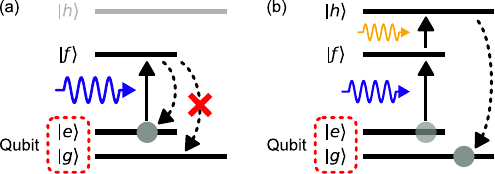}
    \caption{Energy-level diagrams of fluorescence readout and unconditional reset of a fluxonium. Here, $\ket{g}$, $\ket{e}$, $\ket{f}$, and $\ket{h}$ denote the ground, first-excited, second-excited, and third-excited states of the fluxonium, respectively. (a)~Fluorescence readout using the $\ket{e}$--$\ket{f}$ transition. The $\ket{g}$ and $\ket{e}$ states are used as the computational bases. The $\ket{g}$--$\ket{f}$ transition is dipole-forbidden at the sweet spot. (b)~All-microwave unconditional reset. By simultaneously driving the $\ket{e}$--$\ket{f}$ and $\ket{f}$--$\ket{h}$ transitions, the qubit is reset to $\ket{g}$.}
    \label{fig:conceptual}
\end{figure}

In this work, we demonstrate QND fluorescence readout and high-fidelity unconditional reset of a fluxonium at the sweet spot. We implement dissipation engineering to enable fluorescence readout using a filter structure in a planar circuit, which leads to improved integrability and scalability. We perform fluorescence readout using the transition between the first- and second-excited states to form a closed cycle taking advantage of the selection rule of the fluxonium at the sweet spot.  
The unwanted state leakage from the readout subspace during the readout is significantly suppressed, resulting in enhanced QNDness compared to that reported in the previous study~\cite{Cottet2021}. We also achieve fast and high-fidelity unconditional reset of the fluxonium. To this end, we use the high-pass feature of the filter to enable fast decay from a higher excited state to the ground state. 
By simultaneously applying two microwave drives to pump the qubit excitation to the higher excited state that rapidly decays to the ground state, the qubit is reset to the ground state~[Fig.~\ref{fig:conceptual}(b)]. The all-microwave scheme using no perturbative effect is both hardware- and power-efficient.

\section{Dissipation engineering} \label{sec:filter}
The external decay rate from the eigenstates $\ket{i}$ to $\ket{j}$~($\{i,j\}=\{g,e,f,h,\ldots\}$) of a fluxonium via capacitive coupling with a transmission line is given by 
\begin{align}
    \Gamma_{ij}^{\mrm{ext}} \propto \abs{\omega_{ij}}\abs{\braket{i|n|j}}^2\abs{\coth\!\left(\frac{\hbar\omega_{ij}}{2k_{\mrm{B}}T}\right)+1}
    \label{eq:external_decay}
\end{align}
with $\omega_{ij}=\omega_i-\omega_j$, the transition frequency between $\ket{i}$ and $\ket{j}$; $\hat{n}$ is the charge operator of the fluxonium, $\hbar=h/2\pi$ is the reduced Planck constant, $k_{\mrm{B}}$ is the Boltzmann constant, and $T$ is the effective temperature of the environment. Fast readout requires enhancing the decay rate of the readout transition by increasing the coupling to the transmission line. Without an additional design consideration, however, it also increases the external decay rate of the fluxonium’s computational subspace, leading to a short relaxation time of the $\ket{e}$ state. To realize both fast readout and a long lifetime of the qubit, it is necessary to maximize the external decay rate of the readout subspace while minimizing that of the computational subspace. Introducing a filter structure between the fluxonium and transmission line is an effective approach to create a contrast of the decay rate between the transitions. If the coupling between the filter and the transmission line is sufficiently large, the filter can also be treated as part of the external environment for the qubit. When a filter with a power transmittance $T(\omega)$ is inserted between the qubit and the transmission line, the external decay rate is modulated as
\begin{align}
    \Gamma_{ij}^{\mrm{ext}} \propto \abs{T(\omega_{ij})\omega_{ij}}\abs{\bra{i}\hat{n}\ket{j}}^2\abs{\coth\!\left(\frac{\hbar\omega_{ij}}{2k_{\mrm{B}}T}\right)+1}.
    \label{eq:external_decay_filter}
\end{align}
For fast fluorescence readout using the $\ket{e}$--$\ket{f}$ transition and long lifetime of the qubit, $\left|T(\omega)\right|$ must satisfy both $\left|T(\omega_{ef})\right|\sim 1$ and $\left|T(\omega_{ge})\right|\ll 1$. Such dissipation engineering also enables unconditional reset of a fluxonium. Fast unconditional reset can be achieved by transferring the qubit excitation to a higher excited state with a large decay rate to the ground state. This process involves only the qubit's internal degrees of freedom directly coupled to the bath modes in the environment.
 
Here, we design an on-chip filter structure to implement fluorescence readout on a two-dimensional circuit for dissipation engineering. We follow the method proposed in Ref.~\citenum{Yan2023} which enables the design of band-pass filters with any number of stages, center frequency, and bandwidth, by adjusting only the lengths of transmission lines and short-to-ground stubs. We apply the method to a coplanar waveguide~(CPW), the standard transmission line in superconducting quantum circuits. The filter is single-stage~($N=1$) and implemented using a 50-$\Omega$ CPW. The center frequency~($\omega_{\mrm{c}}/2\pi$) and bandwidth~($\Delta\omega/2\pi$) are designed to be 4.6~GHz and 1.0~GHz, respectively. Figure~\ref{fig:device}(d) shows the transmission spectrum of the filter. The filter provides more than 30-dB attenuation below 1~GHz, the typical frequency range for the fluxonium $\ket{g}$--$\ket{e}$ transition. Transmission-line-based filters also transmit signals at integer multiples of the designed center frequency. As shown in Fig.~\ref{fig:device}(d), the filter designed here does not exhibit a simple band-pass characteristic, but rather has a high-pass-like transmission characteristics. We also note that dissipation engineering can be achieved with less footprint by constructing filter circuits using spiral~\cite{Park2024}, lumped-element~\cite{Paik2010, Cho2013, Deng2014, Zotova2023, Zotova2024}, or other compact microwave resonators~\cite{Geerlings2012, Jiang2022} with low external quality factors. Such compact structures are advantageous for integration. The design flexibility of the filters is one of the benefits of using an on-chip structure.

\section{Experiment} \label{sec:experiment}
\subsection{Device} \label{ssec:device}
\begin{figure}[t]
    \centering
    \includegraphics[width=8.6cm, pagebox=cropbox, clip]{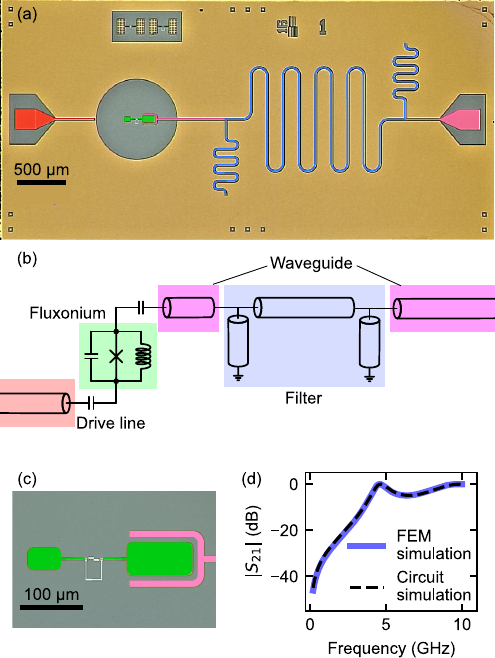}
    \caption{Device studied in this work. (a)~False-colored optical image of the chip. A fluxonium~(green) is capacitively coupled to a coplanar waveguide~(pink), with the CPW filter~(blue) inserted to the middle of the waveguide. Another waveguide~(red) for qubit drive is capacitively coupled to the qubit. (b)~Circuit diagram of the device. (c)~False-colored magnified view of the fluxonium. The qubit and the waveguide are coupled through a U-shaped pad. (d)~Transmission spectra of the filter obtained by finite-element~(FEM) simulation~(solid) using COMSOL~\cite{comsol} and distributed-element circuit simulation~(dashed) using QucsStudio~\cite{qucsstudio}.}
    \label{fig:device}
\end{figure}

Our device comprises a fluxonium and the CPW filter designed in Sec.~\ref{sec:filter}. The fluxonium is coupled to two coplanar waveguides, one for the readout and the other for the control drive. The CPW filter is inserted to the waveguide for the readout. The chip image and the magnified view of the fluxonium are shown in Figs.~\ref{fig:device}(a) and (c), respectively. 

\begin{figure}[tb]
    \centering
    \includegraphics[width=8.6cm, pagebox=cropbox, clip]{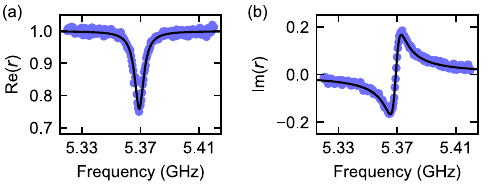}
    \caption{(a)~Real and (b)~imaginary parts of the reflection coefficient $r$ at the sweet spot~(dots). The solid black lines represent the theoretical fits.}
    \label{fig:ef_spectrum}
\end{figure}

Figure~\ref{fig:ef_spectrum} presents the reflection spectrum of the $\ket{e}$--$\ket{f}$ transition measured at the sweet spot. The reflection coefficient is normalized by dividing the measured reflection signal by a background reflection signal at a bias point off the sweet spot. The reflection coefficient fits well to the theoretical prediction,
\begin{align}
    r = 1 - P_{ef}\frac{\Gamma_{\mrm{r}}(i\Delta+\Gamma_{\mrm{r}}/2)}{\Delta^2+\Gamma_{\mrm{r}}^2/4+\Omega^2/2},
    \label{eq:reflection}
\end{align}
where $P_{ef}$ is the population of the $\ket{e}$--$\ket{f}$ subspace, $\Gamma_{\mrm{r}}$ is the external decay rate of the readout transition, $\Delta$ is the detuning between the drive and the $\ket{e}$--$\ket{f}$ transition frequency, and $\Omega$ is the drive amplitude. The derivation of Eq.~(\ref{eq:reflection}) is provided in Appendix~\ref{appx:reflection}. From the fit, we extract the external decay rate of the readout transition as $\Gamma_{\mrm{r}}/2\pi=5.4(1)$~MHz. We also measure the $\ket{g}$--$\ket{e}$ transition frequency as $255$~MHz.

\subsection{QND readout} \label{ssec:readout}
\begin{figure}[t]
    \centering
    \includegraphics[width=8.6cm, pagebox=cropbox, clip]{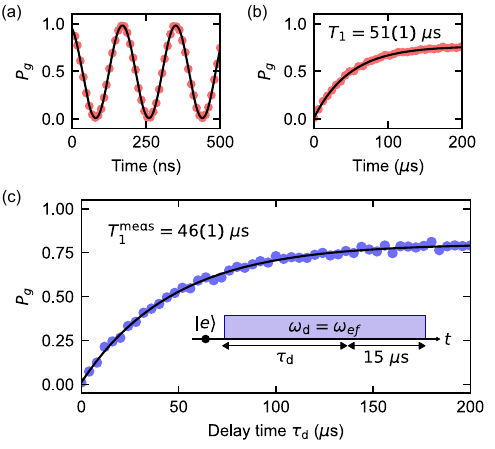}
    \caption{Fluorescence readout using the $\ket{e}$--$\ket{f}$ transition. (a)~Rabi oscillation between the $\ket{g}$ and $\ket{e}$ states. The population of the $\ket{g}$ state, $P_g$, is measured as a function of the qubit control drive pulse. The solid black line shows the fit to a sine function. (b)~Qubit relaxation from the $\ket{e}$ state to the equilibrium state. The solid black line represents the fit to an exponential decay. (c)~Qubit relaxation from the $\ket{e}$ state to the equilibrium state in the presence of the readout drive at $\omega_{\mrm{d}}$. The inset depicts the pulse sequence. The readout signal is integrated for 15~$\mu$s after the delay time $\tau_{\mathrm{d}}$. The solid black curve represents the fit to an exponential decay.}
    \label{fig:time_domain}
\end{figure}

First, we demonstrate fluorescence readout using the $\ket{e}$--$\ket{f}$ transition. We note that the experiments are performed after resetting the qubit into $\ket{g}$ by the scheme explained in Sec.~\ref{ssec:reset}. The reflection signals are converted to the ground-state population $P_g$ by normalization described in Appendix~\ref{appx:singleshot}. 
The readout-pulse length is fixed at 15~$\mu$s, and the readout-pulse amplitude is selected so that the readout signal-to-noise ratio is maximized. We use a lumped-element Josephson parametric amplifier~(JPA) in the non-degenerate mode to amplify the readout signal. Figure~\ref{fig:time_domain}(a) shows a Rabi oscillation between $\ket{g}$ and $\ket{e}$ states. Observation of the Rabi oscillation indicates that the population of the $\ket{e}$ state can be measured via fluorescence readout sufficiently faster than the energy-relaxation time of the computational basis.

 In Fig.~\ref{fig:time_domain}(b), we show the qubit energy relaxation. From the fit to an exponential decay, the energy-relaxation time is extracted as $T_1=51(1)$~$\mu$s. The ratio of the computational subspace lifetime to readout subspace lifetime gives:
\begin{align}
    \frac{T_1}{1/\Gamma_{\mrm{r}}} = 1.7\times 10^3.
\end{align}
The significant contrast between the lifetimes proves the effectiveness of the on-chip filter design. 

On the other hand, Fig.~\ref{fig:time_domain}(c) shows the transient dynamics of the qubit during the fluorescence readout. We prepare the qubit in the $\ket{e}$ state and apply the readout drive for a delay time $\tau_{\mathrm{d}}$ before integrating the readout signal for 15~$\mu$s. The measured ground-state population exhibits an exponential decay as a function of $\tau_{\mrm{d}}$ with a characteristic decay time $T_1^{\mrm{meas}}=46(1)$~$\mu$s. The QNDness of the fluorescence readout is quantified by the average number of fluorescence cycles, $N_{\mrm{QND}}$, that we can drive before the qubit relaxation to the ground state. In our case, $N_{\mrm{QND}}$ is calculated as
\begin{align}
N_{\mrm{QND}}=\Gamma_{\mrm{r}}T_1^{\mrm{meas}}=1.6\times 10^3.
\end{align}
In the previous study~\cite{Cottet2021}, the characteristic decay time during the readout was $T_1^{\mrm{meas}}=9.6$~$\mu$s and the QNDness of the fluorescence readout was $N_{\mrm{QND}}=1.1\times 10^2$. The use of the $\ket{e}$--$\ket{f}$ transition instead of the $\ket{g}$--$\ket{h}$ transition enables an order-of-magnitude improvement in QNDness compared to that of the previous work. 

To further investigate the non-QNDness of our fluorescence readout, we calculate the characteristic decay time purely induced by the measurement as
\begin{align}
    \left(\frac{1}{T_1^{\mrm{meas}}} - \frac{1}{T_1}\right)^{-1} = 0.55~\text{ms}.
\end{align}
This value approaches infinity if the QNDness of the readout is limited by $T_1$. We attribute the finite value observed in our experiment to the quasiparticle tunneling in the single Josephson junction. Although the $\ket{g}$--$\ket{f}$ transition is dipole-forbidden at the sweet spot, the operator associated with the energy relaxation due to the quasiparticle tunneling in the single Josephson junction, $\sin\!\left(\frac{\hat{\phi}+\phi_{\mrm{ext}}}{2}\right)$, has a non-zero matrix element between $\ket{g}$ and $\ket{f}$ at the sweet spot. The measured $T_1^{\mrm{meas}}$ is explained by assuming the quasiparticle density around the single Josephson junction $x_{\mrm{qp}}=4\times 10^{-7}$, which is within the typical range observed in experiments~\cite{Martinis2009, Vool2014, Pop2014, Serniak2018, Nguyen2019}. Detailed discussion can be found in Appendix~\ref{appx:pump}. We can achieve better QNDness by suppressing quasiparticle tunneling through careful filtering of high-frequency noise~\cite{Connolly2024} and other techniques such as superconducting gap engineering~\cite{Aumentado2004, Kalashnikov2020, Connolly2024, McEwen2024}. 

Compared to dispersive readout, the fluorescence readout of fluxoniums offers significantly reduced modeling and analysis complexity. As discussed in previous studies~\cite{Shillito2022, Dumas2024}, analyzing the dynamics of the qubit--resonator system in detail requires a substantial computational cost. This complexity makes it challenging to identify and model the mechanism that degrade the QNDness in dispersive readout of fluxoniums. In contrast, fluorescence readout involves only the qubit's degree of freedom, significantly simplifying the modeling and analysis~\cite{Cottet2021}. 

To further speed up the readout, it is effective to enhance the coupling between the waveguide and the qubit, thereby increasing $\Gamma_{\mrm{r}}$. Since $\Gamma_{\mrm{r}}$ is proportional to the square of the coupling capacitance between the waveguide and the qubit~\cite{Moskalenko2022}, increasing the coupling capacitance by a factor of 10 through optimization of the geometry causes $\Gamma_{\mrm{r}}$ to increase by a factor of 100. This may help achieve a readout speed comparable to the current state of the art~\cite{Walter2017, Sunada2022,Swiadek2024,Sunada2024,Spring2024} using fluorescence readout.

\subsection{Unconditional reset} \label{ssec:reset}
We demonstrate an all-microwave active unconditional reset leveraging the engineered dissipation. The $\ket{g}$--$\ket{h}$ transition frequency of the qubit~(7.814~GHz) lies within the passband of the filter. The external decay rate from $\ket{h}$ to $\ket{g}$ is estimated as $\Gamma_{hg}/2\pi=6.5$~MHz using Eq.~(\ref{eq:external_decay_filter}) and the measured $\Gamma_{\mrm{r}}$. By simultaneously driving the $\ket{e}$--$\ket{f}$ and $\ket{f}$--$\ket{h}$ transitions, any excitation in the \{$\ket{e}$, $\ket{f}$, $\ket{h}$\} subspace is rapidly and unconditionally reset to the $\ket{g}$ state~[Fig.~\ref{fig:conceptual}(b)]. In Appendix~\ref{appx:calib}, we explain the optimization of the frequencies and amplitudes of the $\ket{e}$--$\ket{f}$ and $\ket{f}$--$\ket{h}$ drives.

\begin{figure}[t]
    \centering
    \includegraphics[width=8.6cm, pagebox=cropbox, clip]{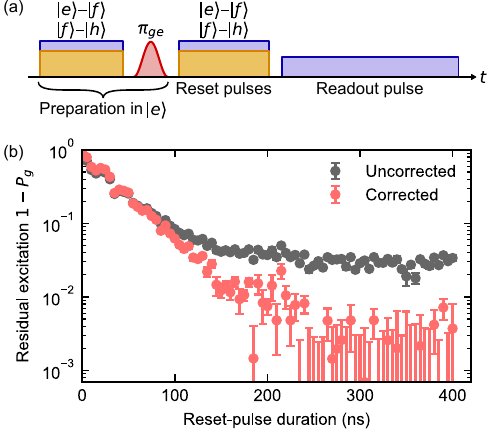}
    \caption{Unconditional reset of the fluxonium. (a)~Pulse sequence for evaluating the reset fidelity. (b)~Measured residual excitation $1-P_g$ as a function of the reset-pulse duration. The gray dots represent the raw data, while the red dots depict the data corrected for the excitation during the readout pulse. The error bars are obtained using Eq.~(\ref{appxeq:Pg_correction}) and error propagation.}
    \label{fig:reset}
\end{figure}

We use the pulse sequence shown in Fig.~\ref{fig:reset}(a) to demonstrate the unconditional reset. The qubit is prepared in $\ket{e}$ by calibrated 1-$\mu$s reset pulses, followed by a $\pi$ pulse between $\ket{g}$ and $\ket{e}$. The readout-pulse length to measure the residual excitation is fixed at 15~$\mu$s. As discussed in Sec.~\ref{ssec:readout}, the ground-state population under the readout drive follows an exponential decay. The measured residual excitation is the average value over the readout-pulse duration. The true residual population just before the readout pulse is then obtained by correcting the excitation of the qubit during the readout. The correction method is presented in detail in Appendix~\ref{appx:correction}. 

Figure~\ref{fig:reset}(b) shows the residual excitation of the qubit as a function of the reset-drive length. The residual excitation for pulse lengths of \{50, 100, 150, 200\} ns are \{25.4(3), 8.4(4), 4.1(2), 3.7(1)\}\% before correcting for the excitation during the readout pulse and \{26.2(3), 6.2(4), 1.2(3), 0.7(2)\}\% after the correction. Based on the corrected values, the residual excitation falls below 1\% for pulse lengths $> 200$~ns and reaches a level comparable to statistical errors~($\sim$0.3\%) for pulse lengths $> 250$~ns. The theoretical limit for the reset speed is the external decay rate from $\ket{h}$ to $\ket{g}$. We may further improve the reset speed by fine optimization of the reset-pulse waveforms via simultaneous sweep of the frequencies and amplitudes or machine-learning-assisted optimization~\cite{Ding2023,Li2024,Chatterjeee2024}.

Our method is all-microwave and does not utilize a fast flux-bias pulse. Moreover, one of the two microwave drives used in the reset pulse has the same frequency as that of the readout drive, minimizing the additional hardware resources required. Another advantage of our method is that we only use the qubit degree of freedom, and no perturbation is employed. As a result, the required microwave power is reduced by more than two orders of magnitude compared to that of existing all-microwave schemes~\cite{Magnard2018, Sunada2022, Zhang2021}.

\section{Conclusions} \label{sec:discussion}
We demonstrated a non-demolition fluorescence readout and a high-fidelity unconditional reset of a fluxonium via an on-chip dissipation engineering. We designed a coplanar-waveguide filter that protects the computational subspace from decay while maintaining the external decay rate of the readout transition. Taking advantage of the engineered dissipation, we realized fluorescence readout with an order-of-magnitude improvement in the QNDness, compared to the previous study, by choosing the $\ket{e}$--$\ket{f}$ transition as the readout transition. We attributed the non-QNDness of our shelving readout to the quasiparticle tunneling in the single Josephson junction. Techniques such as superconducting gap engineering can contribute to further improving the QNDness. Furthermore, we proposed and demonstrated fast, high-fidelity, and all-microwave unconditional reset of the fluxonium via dissipation engineering. We achieved over 99\% reset fidelity with a duration of 200~ns, and above 99.5\% reset fidelity with a duration of 250~ns. Our unconditional reset scheme is both hardware- and power-efficient with no requirement for a fast flux-bias pulse or a strong microwave drive of a higher-order transition. A sophisticated optimization of the reset-pulse waveforms can be effective for realizing even faster unconditional reset. Our results pave the way for superconducting quantum computing free from dispersive interaction between qubits and resonators, offering new directions for architectural design.

\section{Acknowledgements}
We thank K. Kato and T. Miyamura for fruitful discussions. The work was supported in part by MEXT Q-LEAP~(Grant No.~JPMXS0118068682), JSPS Grant-in-Aid for Scientific Research~(KAKENHI)~(Grant No.~JP22H04937),  WINGS-QSTEP by the University of Tokyo, JST SPRING~(Grant No.~JPMJSP2108), and JST CREST~(Grant No.~JPMJCR23I4).

\appendix
\renewcommand{\thefigure}{\thesection\arabic{figure}}
\section{Experimental setup} \label{appx:setup}
\setcounter{figure}{0}
The measurement setup used in this work is shown in Fig.~\ref{fig:wiring}. The $\ket{g}$--$\ket{e}$ drive is directly generated by an arbitrary waveform generator~(AWG), while the $\ket{e}$--$\ket{f}$ and $\ket{f}$--$\ket{h}$ drives are generated by up-converting the signals from dedicated AWGs with continuous-wave drives from signal generators using single-sideband mixers. We use a JPA in the non-degenerate mode to amplify the readout signal. The pump drive for the JPA is directly input from a signal generator and always on during the readout and reset experiment.  

The parameters of the fluxonium used in this work are shown in Table~\ref{tab:parameter}. 

\begin{figure}[t]
    \centering
    \includegraphics[width=8.6cm, pagebox=cropbox, clip]{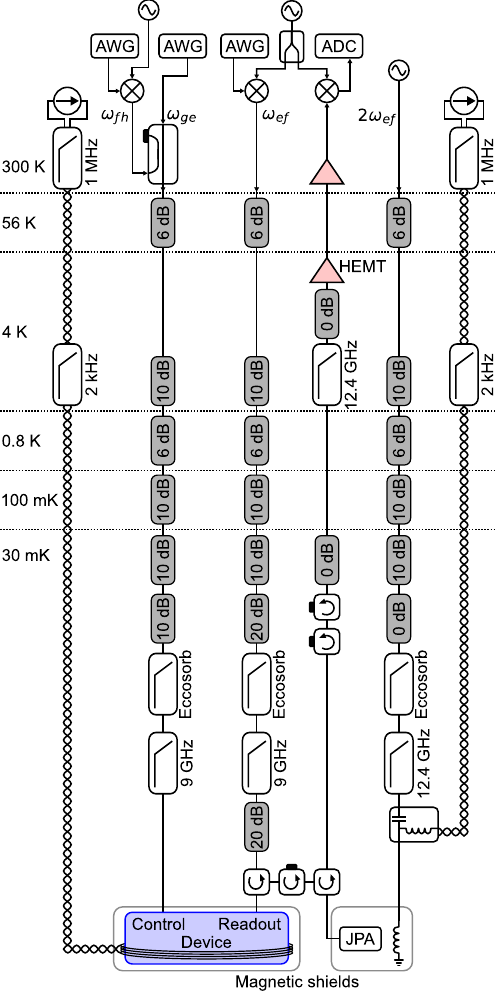}
    \caption{Measurement setup. AWG, arbitrary-waveform generator; ADC, analog-to-digital converter; HEMT, high-electron-mobility transistor; JPA, Josephson parametric amplifier.}
    \label{fig:wiring}
\end{figure}

\begin{table}[b]
    \centering
    \caption{Parameters of the fluxonium.}
    \begin{tabular}{lcr}
    \hline\hline
        Josephson energy & $E_{\mrm{J}}/h$ & 6.23~GHz\\
        Charging energy & $E_{\mrm{C}}/h$ & 1.25~GHz\\
        Inductive energy & $E_{\mrm{L}}/h$ & 0.86~GHz\\\hline
        $\ket{g}$--$\ket{e}$ transition frequency&$\omega_{ge}/2\pi$& 255~MHz \\
        $\ket{e}$--$\ket{f}$ transition frequency&$\omega_{ef}/2\pi$& 5.369~GHz \\
        $\ket{f}$--$\ket{h}$ transition frequency&$\omega_{fh}/2\pi$& 2.215~GHz \\
        $\ket{g}$--$\ket{h}$ transition frequency&$\omega_{gh}/2\pi$& 7.814~GHz \\\hline
        Energy-relaxation time& $T_1$ & $51(1)$~$\mu$s \\
        Ramsey dephasing time& $T_2^*$& $32.9(6)$~$\mu$s \\
        Echo dephasing time& $T_2^{\mrm{echo}}$ & $33(1)$~$\mu$s \\\hline
        Qubit thermal population & $P_g^{\mrm{th}}$ & $0.77(1)$\\
        Single-qubit Clifford-gate fidelity & $F_{\mrm{1Q}}$ & $99.940(5)$\%\\
    \hline\hline
    \end{tabular}
    \label{tab:parameter}
\end{table}

\section{Fluorescence readout} \label{appx:readout}
\setcounter{figure}{0}
\subsection{Reflection coefficient} \label{appx:reflection}
We consider a fluxonium system up to the second-excited state~($\{\ket{g},\ket{e},\ket{f}\}$). The Hamiltonian of the driven system with the drive frequency $\omega_{\mrm{d}}$ and amplitude $\Omega$ is given by
\begin{align}
    \begin{split}
        \hat{H}/\hbar =&\, \omega_{ge}\ketbra{e}{e} + (\omega_{ge}+\omega_{ef})\ketbra{f}{f} \\+&\, \Omega\cos\omega_{\mrm{d}}t\left(\eta\ketbra{e}{g} + \ketbra{f}{e} + \mrm{h.c.}\right),
    \end{split}
\end{align}
where $\omega_{ge}$ and $\omega_{ef}$ are the $\ket{g}$--$\ket{e}$ and $\ket{e}$--$\ket{f}$ transition frequencies, respectively, and $\eta=\abs{\bra{e}\hat{n}\ket{g}/\bra{f}\hat{n}\ket{e}}$ is the ratio of the charge matrix elements between the relevant states. Moving onto a rotating frame by a unitary transformation $\hat{U}=\exp\left\{-it\left[\omega_{ge}\ketbra{e}{e} + (\omega_{ge}+\omega_{\mrm{d}})\ketbra{f}{f}\right]\right\}$ and applying the rotating-wave approximation yield 
\begin{align}
    \hat{H}'/\hbar = \Delta\ketbra{f}{f} + \frac{\Omega}{2} \left(\ketbra{f}{e} + \ketbra{e}{f}\right),
\end{align}
where $\Delta=\omega_{ef}-\omega_{\mrm{d}}$ is the detuning between the drive and $\ket{e}$--$\ket{f}$ transition frequencies, and we assume $\abs{\omega_{ef}-\omega_{\mrm{d}}}\ll\abs{\omega_{ge}-\omega_{\mrm{d}}}$. Assuming that the external decay to the transmission line is the dominant decay channel for the $\ket{e}$--$\ket{f}$ transition and that the temperature is sufficiently low compared to the transition frequency~($\hbar\omega_{ef}\gg k_{\mrm{B}}T$), the master equation is written as
\begin{align}
    \dif{\hat{\rho}}{t} = -\frac{i}{\hbar}\left[\hat{H}',\hat{\rho}\right] + \Gamma_{\mrm{r}}\hat{\mcal{D}}(\ketbra{e}{f})\hat{\rho},
\end{align}
where $\hat{\mcal{D}}$ is the Lindblad superoperator. The equations of motion are written as 
\begin{align}
    \begin{cases}
        \dot{\rho}_{ee} = -i\frac{\Omega}{2}(\rho_{fe}-\rho_{ef}) + \Gamma_{\mrm{r}}\rho_{ff},\\
        \dot{\rho}_{ff} = i\frac{\Omega}{2}(\rho_{fe}-\rho_{ef}) - \Gamma_{\mrm{r}}\rho_{ff},\\
        \dot{\rho}_{ef} = -i\frac{\Omega}{2}(\rho_{ff}-\rho_{ee}) + \left(i\Delta-\frac{\Gamma_{\mrm{r}}}{2}\right)\rho_{ef},\\
        \dot{\rho}_{fe} = i\frac{\Omega}{2}(\rho_{ff}-\rho_{ee}) - \left(i\Delta+\frac{\Gamma_{\mrm{r}}}{2}\right)\rho_{fe}.
    \end{cases}
\end{align}
The diagonal elements of the density matrix satisfy the conservation law $\rho_{ee}+\rho_{ff}=P_{ef}$ , where $P_{ef}$ is the population of the $\ket{e}$--$\ket{f}$ subspace. The steady-state solutions of the equations are calculated as
\begin{align}
    \begin{cases}
        \rho_{ee} = P_{ef}\dfrac{\Delta^2+\Gamma_{\mrm{r}}^2/4+\Omega^2/4}{\Delta^2+\Gamma_{\mrm{r}}^2/4+\Omega^2/2},\\
        \rho_{ff} = P_{ef}\dfrac{\Omega^2/4}{\Delta^2+\Gamma_{\mrm{r}}^2/4+\Omega^2/2},\\
        \rho_{ef} = -iP_{ef}\dfrac{\Omega}{2}\dfrac{i\Delta+\Gamma_{\mrm{r}}/2}{\Delta^2+\Gamma_{\mrm{r}}^2/4+\Omega^2/2},\\
        \rho_{fe} = \rho_{ef}^{*}.
    \end{cases}
\end{align}

The input--output relation is written as $\hat{b}_{\mrm{out}} = \hat{b}_{\mrm{in}} + \sqrt{\Gamma_{\mrm{r}}}\,\ketbra{e}{f}$, where $\hat{b}_{\mrm{in}}$~($\hat{b}_{\mrm{out}}$) is the operator representing the input~(output) signal. Considering the steady state and replacing the input~(output) operator $\hat{b}_{\mrm{in}}$~($\hat{b}_{\mrm{out}}$) with the expectation values $\beta_{\mrm{in}}=\braket{\hat{b}_{\mrm{in}}}$~($\beta_{\mrm{out}}=\braket{\hat{b}_{\mrm{out}}}$), the input--output relation is rewritten as $\beta_{\mrm{out}} = \beta_{\mrm{in}} + \sqrt{\Gamma_{\mrm{r}}}\,\rho_{ef}$. Using the relation $\Omega = -2i\sqrt{\Gamma_{\mrm{r}}}\,\beta_{\mrm{in}}$, we obtain
\begin{align}
    r = \frac{\beta_{\mrm{out}}}{\beta_{\mrm{in}}} = 1 - P_{ef}\frac{\Gamma_{\mrm{r}}(i\Delta+\Gamma_{\mrm{r}}/2)}{\Delta^2+\Gamma_{\mrm{r}}^2/4+\Omega^2/2}.
    \label{appxeq:reflection}
\end{align}
Equation~(\ref{appxeq:reflection}) indicates that the reflection coefficient changes depending on the population of the $\ket{e}$--$\ket{f}$ subspace. Therefore, measuring the reflection coefficient enables the fluorescence readout of the qubit states. 

The signal-to-noise ratio~(SNR) of the fluorescence readout is maximized when the difference  $\abs{\beta_{\mrm{out}}^g-\beta_{\mrm{out}}^e}$ is maximized, with $\beta_{\mrm{out}}^g$~($\beta_{\mrm{out}}^e$) being the complex amplitude of the output signal for the qubit in the $\ket{g}$~($\ket{e}$) state. Using the input--output relation, the distance $\abs{\beta_{\mrm{out}}^g-\beta_{\mrm{out}}^e}$ is calculated as
\begin{align}
    \begin{split}
        \abs{\beta_{\mrm{out}}^g-\beta_{\mrm{out}}^e}^2 &= \Gamma_{\mrm{r}}\abs{\rho_{ef}(P_{ef}=0) - \rho_{ef}(P_{ef}=1)}^2\\
        &= \frac{\Gamma_{\mrm{r}}\Omega^2}{4}\frac{\Delta^2+\Gamma_{\mrm{r}}^2/4}{\left(\Delta^2+\Gamma_{\mrm{r}}^2/4+\Omega^2/2\right)^2},
    \end{split}
\end{align}
which is maximized at $\Omega=\Gamma_{\mrm{r}}/\sqrt{2}$ when the detuning $\Delta$ is zero. At the optimal readout condition, the reflection coefficient $r$ reduces to
\begin{align}
    r = 1-P_{ef}.
\end{align}

\subsection{Readout signals and state populations} \label{appx:singleshot}
\begin{figure}[t]
    \centering
    \includegraphics[width=8.6cm, pagebox=cropbox, clip]{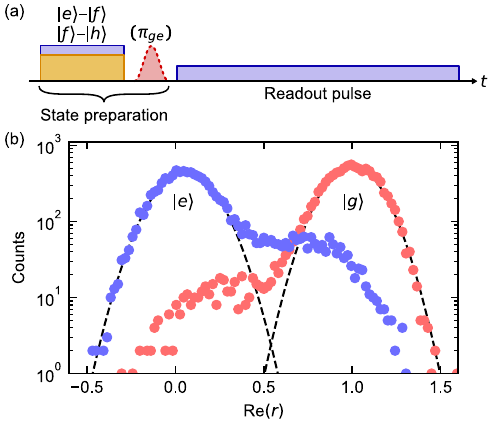}
    \caption{Single-shot fluorescence readout of the fluxonium. (a)~Pulse sequence for the single-shot readout experiment. The initialization pulse is 1-$\mu$s long with optimized amplitudes of the two-tone drives. The readout pulse duration is 15~$\mu$s. The sequence is repeated every 200~$\mu$s. (b)~Histograms of time-integrated single-shot reflection coefficients projected on the real axis. The total shot number is 10000. Dashed lines represent the fits of the $\ket{g}$- and $\ket{e}$-state signals to the Gaussian distributions. }
    \label{fig:histogram}
\end{figure}

Figure~\ref{fig:histogram}(b) shows the histograms of time-integrated single-shot reflection coefficients. As derived from Eq.~(\ref{appxeq:reflection}), the reflection coefficients take real values when the detuning $\Delta$ is zero. Therefore, we project the reflection coefficient onto the real axis to discriminate the outcome of the readout. From the fits to the Gaussian distributions, the SNR of the measurement is calculated as
\begin{align}
    \text{SNR} = \frac{\abs{\mu_g-\mu_e}}{(\sigma_g+\sigma_e)/2}=6.3(1),
\end{align}
where $\mu_{g(e)}$ and $\sigma_{g(e)}$ are the mean and standard deviation of the $\ket{g}$-($\ket{e}$-)state distribution, respectively. 

We emphasize that the fact that the mean of the $\ket{e}$-state distribution is located near zero indicates that our readout condition is in the vicinity of the optimal readout condition $\Omega=\Gamma_{\mrm{r}}/\sqrt{2}$.

It should also be noted that the histograms deviate from the sum of two Gaussian distributions. We attribute this discrepancy to the saturation of the JPA and the state transition during the readout. When qubit-state transitions occur during readout, outcomes of time-integrated signals are not binary; instead, they exhibit tails between two peaks. In such cases, evaluation of state populations by classifying each shot into either $\ket{g}$ or $\ket{e}$ is not accurate. Instead, evaluation of the state populations should be based on ensemble-averaged values of readout signals for each pulse. For this reason, while the SNR of the single-shot measurement is sufficiently good, we measure the ground-state population of the qubit based on the ensemble-averaged values of the time-integrated reflection coefficients. We calculate the ground-state population by 
\begin{align}
    P_g = \mrm{Re}\left(\frac{r-\mu_e}{\mu_g-\mu_e}\right),
\end{align}
where $r$ is the averaged reflection coefficient. We further correct the state transition during the readout by the method presented in Appendix~\ref{appx:correction}.
 
\subsection{Rate-equation model} \label{appx:pump}
Here, we analyze the non-QNDness of our readout observed in Sec.~\ref{ssec:readout} with a rate-equation model. We assume that the qubit population changes slowly compared to $\Gamma_{\mrm{r}}$, allowing us to consider that the $\ket{e}$--$\ket{f}$ subspace stays in the steady state and that the ratio $\rho_{ee}/\rho_{ff}$ remains constant:
\begin{align}
    \frac{\rho_{ee}}{\rho_{ff}} = \frac{P_e}{P_f} = \left(\frac{\Gamma_{\mrm{r}}}{\Omega}\right)^2 + 1,
\end{align}
where we also assume $\Delta=0$. This approximation gives the population relations
\begin{align}
    \begin{cases}
        P_e = \dfrac{\Gamma_{\mrm{r}}^2+\Omega^2}{\Gamma_{\mrm{r}}^2+2\Omega^2}\left(1-P_g\right),\\
        P_f = \dfrac{\Omega^2}{\Gamma_{\mrm{r}}^2+2\Omega^2}\left(1-P_g\right).
    \end{cases}
\end{align}
The rate equation for the ground-state population under irradiation of the readout drive is given by
\begin{align}
\begin{split}
    \dot{P}_g =& -\Gamma_{ge}P_g + \Gamma_{eg}P_e + \Gamma_{fg}P_f\\
    =& -\Gamma_{ge}P_g\\ &+ \Gamma_{eg}\dfrac{\Gamma_{\mrm{r}}^2+\Omega^2}{\Gamma_{\mrm{r}}^2+2\Omega^2}\left(1-P_g\right) + \Gamma_{fg}\dfrac{\Omega^2}{\Gamma_{\mrm{r}}^2+2\Omega^2}\left(1-P_g\right)\\
    =& -\left[\Gamma_{ge} + \Gamma_{eg} + \left(\Gamma_{fg}-\Gamma_{eg}\right)\dfrac{\Omega^2}{\Gamma_{\mrm{r}}^2+2\Omega^2}\right]P_g + \mrm{const.},
\end{split}
\end{align}
where $\Gamma_{ij}$ is the decay rate from $\ket{i}$ to $\ket{j}$. Therefore, the effective relaxation rate during the readout is 
\begin{align}
    \Gamma_1^{\mrm{meas}} = \Gamma_{ge} + \Gamma_{eg} + \left(\Gamma_{fg}-\Gamma_{eg}\right)\dfrac{\Omega^2}{\Gamma_{\mrm{r}}^2+2\Omega^2}.
    \label{eq:gamma1_meas}
\end{align}
We assume that $\Gamma_{ge}$, $\Gamma_{eg}$, and $\Gamma_{fg}$ are not affected by the readout drive. As $\Gamma_{ge} + \Gamma_{eg}=\Gamma_1$ represents the ordinary qubit energy-relaxation rate, the remaining component corresponds to the relaxation rate induced purely by the readout drive. Using $\Gamma_{eg}$ estimated from the $T_1$ decay curve and assuming $\Omega=\Gamma_{\mrm{r}}/\sqrt{2}$, $1/\Gamma_{fg}$ is estimated to be $41.6$~$\mu$s. As discussed in Sec.~\ref{ssec:readout}, we attribute the origin of $\Gamma_{fg}$ to the quasiparticle tunneling in the single Josephson junction of the fluxonium. The energy-relaxation rate due to the quasiparticle tunneling in the single Josephson junction is given by~\cite{Koch2011}
\begin{align}
    \Gamma_{ij}^{\mrm{qp}} = \frac{8E_{\mrm{J}}}{\pi\hbar}x_{\mrm{qp}}\sqrt{\frac{2\Delta_{\mrm{SC}}}{\hbar\omega_{ij}}}\abs{\braket{i|\sin\!\left(\frac{\hat{\phi}+\phi_{\mrm{ext}}}{2}\right)\!|j}}^2,
    \label{appxeq:quasiparticle}
\end{align}
where $x_{\mrm{qp}}$ is the quasiparticle density and $\Delta_{\mrm{SC}}$ is the superconducting gap. Using Eq.~(\ref{appxeq:quasiparticle}), we estimate $x_{\mrm{qp}}=4\times 10^{-7}$, which is within the typical range observed in experiments~\cite{Martinis2009, Vool2014, Pop2014, Serniak2018, Nguyen2019}. We note that the estimated value of $x_{\mrm{qp}}$ is too large to account for the relaxation rate of the computational subspace: The measured $T_1$ sets an upper bound of $x_{\mrm{qp}} < 5\times 10^{-8}$ in the Josephson junction array. Although the physical mechanism remains unclear, it has been reported that the quasiparticle density around a single Josephson junction is more than an order of magnitude larger than that in a Josephson junction array~\cite{Somoroff2023}. 

Equation~(\ref{eq:gamma1_meas}) suggests that when $\Gamma_{fg}<\Gamma_{eg}$, the relaxation time during the readout is longer than the ordinary energy-relaxation time. This implies that the QNDness limit of fluorescence readout can exceed the $T_1$ limit.

\section{Calibration of unconditional reset} \label{appx:calib}
\setcounter{figure}{0}
\begin{figure}[t]
    \centering
    \includegraphics[width=8.6cm, pagebox=cropbox, clip]{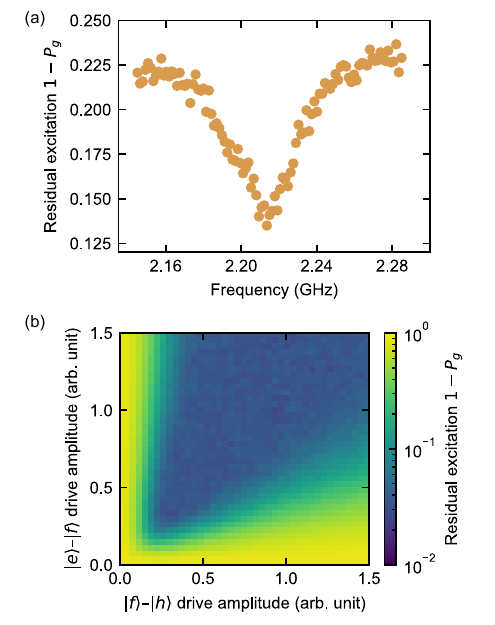}
    \caption{Calibration of the reset protocol. (a)~Spectroscopy of the $\ket{f}$--$\ket{h}$ transition. (b)~Optimization of the $\ket{e}$--$\ket{f}$ and $\ket{f}$--$\ket{h}$ drive amplitudes.}
    \label{fig:calibration}
\end{figure}
Here, we explain the calibration protocol of our unconditional reset. We first determine the $\ket{f}$--$\ket{h}$ transition frequency by sweeping the drive frequency and measuring the ground-state population while applying the $\ket{e}$--$\ket{f}$ drive. At the resonance, the $\ket{e}$ population is excited to $\ket{h}$ by the two drives and subsequently relaxes to $\ket{g}$, resulting in the observed residual excitation reaching the minimum. The result is shown in Fig.~\ref{fig:calibration}(a). 

We next optimize the amplitudes of the $\ket{e}$--$\ket{f}$ and $\ket{f}$--$\ket{h}$ drives. Figure~\ref{fig:calibration}(b) shows the changes in the ground-state population as functions of the drive amplitudes. We select the amplitudes where the residual excitation is the minimum. We note that the optimal $\ket{e}$--$\ket{f}$ drive amplitude is an order-of-magnitude larger than the readout drive.

\section{Correction for state transition during readout} \label{appx:correction}
\setcounter{figure}{0}
As discussed in Sec.~\ref{ssec:readout}, the ground-state population under the irradiation of the readout drive follows an exponential decay. The measured values of the ground-state population are different from the true values right before the readout pulse. We correct this state transition by a method similar to Ref.~\citenum{Somoroff2023}. The time evolution of the ground-state population during the readout pulse follows the exponential decay:
\begin{align}
    P_{g}(t) = P_{g}(\infty) + \left(P_{g}(0) - P_{g}(\infty)\right)e^{-t/T_1^{\mrm{meas}}}.
\end{align}
Assuming that the measured ground-state population $\tilde{P}_{g}$ is the average value over the readout pulse duration $\tau$, $\tilde{P}_{g}$ is given by 
\begin{align}
    \begin{split}
    \tilde{P}_{g} =&\, \frac{1}{\tau}\int_0^{\tau}\mrm{d}t\, P_{g}(t)\\
     =&\, \frac{T_1^{\mrm{meas}}}{\tau}\left(P_{g}(0) - P_{g}(\infty)\right)\left(1-e^{-\tau/T_1^{\mrm{meas}}}\right) \\
     &+  P_{g}(\infty).
    \end{split}
    \label{appxeq:Pg_measure}
\end{align}
The ground-state population right before the readout pulse $P_{g}(0)$ is obtained by solving Eq.~(\ref{appxeq:Pg_measure}) for $P_{g}(0)$:
\begin{align}
    P_{g}(0) = P_{g}(\infty) + \frac{\tau}{T_1^{\mrm{meas}}}\frac{\tilde{P}_{g}-P_{g}(\infty)}{1-e^{-\tau/T_1^{\mrm{meas}}}}.
    \label{appxeq:Pg_correction}
\end{align}
The relaxation time during the readout, $T_1^{\mrm{meas}}$, is obtained in Sec.~\ref{ssec:readout}. We also determine $P_{g}(\infty)$ to be $0.794(5)$ by measuring the steady-state value under the readout drive. 

\bibliography{main}

\begin{thebibliography}{72}%
\makeatletter
\providecommand \@ifxundefined [1]{%
 \@ifx{#1\undefined}
}%
\providecommand \@ifnum [1]{%
 \ifnum #1\expandafter \@firstoftwo
 \else \expandafter \@secondoftwo
 \fi
}%
\providecommand \@ifx [1]{%
 \ifx #1\expandafter \@firstoftwo
 \else \expandafter \@secondoftwo
 \fi
}%
\providecommand \natexlab [1]{#1}%
\providecommand \enquote  [1]{``#1''}%
\providecommand \bibnamefont  [1]{#1}%
\providecommand \bibfnamefont [1]{#1}%
\providecommand \citenamefont [1]{#1}%
\providecommand \href@noop [0]{\@secondoftwo}%
\providecommand \href [0]{\begingroup \@sanitize@url \@href}%
\providecommand \@href[1]{\@@startlink{#1}\@@href}%
\providecommand \@@href[1]{\endgroup#1\@@endlink}%
\providecommand \@sanitize@url [0]{\catcode `\\12\catcode `\$12\catcode `\&12\catcode `\#12\catcode `\^12\catcode `\_12\catcode `\%12\relax}%
\providecommand \@@startlink[1]{}%
\providecommand \@@endlink[0]{}%
\providecommand \url  [0]{\begingroup\@sanitize@url \@url }%
\providecommand \@url [1]{\endgroup\@href {#1}{\urlprefix }}%
\providecommand \urlprefix  [0]{URL }%
\providecommand \Eprint [0]{\href }%
\providecommand \doibase [0]{https://doi.org/}%
\providecommand \selectlanguage [0]{\@gobble}%
\providecommand \bibinfo  [0]{\@secondoftwo}%
\providecommand \bibfield  [0]{\@secondoftwo}%
\providecommand \translation [1]{[#1]}%
\providecommand \BibitemOpen [0]{}%
\providecommand \bibitemStop [0]{}%
\providecommand \bibitemNoStop [0]{.\EOS\space}%
\providecommand \EOS [0]{\spacefactor3000\relax}%
\providecommand \BibitemShut  [1]{\csname bibitem#1\endcsname}%
\let\auto@bib@innerbib\@empty
\bibitem [{\citenamefont {Kjaergaard}\ \emph {et~al.}(2020)\citenamefont {Kjaergaard}, \citenamefont {Schwartz}, \citenamefont {Braumüller}, \citenamefont {Krantz}, \citenamefont {Wang}, \citenamefont {Gustavsson},\ and\ \citenamefont {Oliver}}]{Kjaergaard2020}%
  \BibitemOpen
  \bibfield  {author} {\bibinfo {author} {\bibfnamefont {M.}~\bibnamefont {Kjaergaard}}, \bibinfo {author} {\bibfnamefont {M.~E.}\ \bibnamefont {Schwartz}}, \bibinfo {author} {\bibfnamefont {J.}~\bibnamefont {Braumüller}}, \bibinfo {author} {\bibfnamefont {P.}~\bibnamefont {Krantz}}, \bibinfo {author} {\bibfnamefont {J.~I.-J.}\ \bibnamefont {Wang}}, \bibinfo {author} {\bibfnamefont {S.}~\bibnamefont {Gustavsson}},\ and\ \bibinfo {author} {\bibfnamefont {W.~D.}\ \bibnamefont {Oliver}},\ }\href {https://doi.org/https://doi.org/10.1146/annurev-conmatphys-031119-050605} {\bibfield  {journal} {\bibinfo  {journal} {Annu. Rev. Condens. Matter Phys.}\ }\textbf {\bibinfo {volume} {11}},\ \bibinfo {pages} {369} (\bibinfo {year} {2020})}\BibitemShut {NoStop}%
\bibitem [{\citenamefont {Bravyi}\ \emph {et~al.}(2022)\citenamefont {Bravyi}, \citenamefont {Dial}, \citenamefont {Gambetta}, \citenamefont {Gil},\ and\ \citenamefont {Nazario}}]{Bravyi2022}%
  \BibitemOpen
  \bibfield  {author} {\bibinfo {author} {\bibfnamefont {S.}~\bibnamefont {Bravyi}}, \bibinfo {author} {\bibfnamefont {O.}~\bibnamefont {Dial}}, \bibinfo {author} {\bibfnamefont {J.~M.}\ \bibnamefont {Gambetta}}, \bibinfo {author} {\bibfnamefont {D.}~\bibnamefont {Gil}},\ and\ \bibinfo {author} {\bibfnamefont {Z.}~\bibnamefont {Nazario}},\ }\href {https://doi.org/https://doi.org/10.1063/5.0082975} {\bibfield  {journal} {\bibinfo  {journal} {J. Appl. Phys.}\ }\textbf {\bibinfo {volume} {132}},\ \bibinfo {pages} {160902} (\bibinfo {year} {2022})}\BibitemShut {NoStop}%
\bibitem [{\citenamefont {Nakamura}\ \emph {et~al.}(1999)\citenamefont {Nakamura}, \citenamefont {Pashkin},\ and\ \citenamefont {Tsai}}]{Nakamura1999}%
  \BibitemOpen
  \bibfield  {author} {\bibinfo {author} {\bibfnamefont {Y.}~\bibnamefont {Nakamura}}, \bibinfo {author} {\bibfnamefont {Y.~A.}\ \bibnamefont {Pashkin}},\ and\ \bibinfo {author} {\bibfnamefont {J.~S.}\ \bibnamefont {Tsai}},\ }\href {https://doi.org/https://doi.org/10.1038/19718} {\bibfield  {journal} {\bibinfo  {journal} {Nature}\ }\textbf {\bibinfo {volume} {398}},\ \bibinfo {pages} {786} (\bibinfo {year} {1999})}\BibitemShut {NoStop}%
\bibitem [{\citenamefont {Mooij}\ \emph {et~al.}(1999)\citenamefont {Mooij}, \citenamefont {Orlando}, \citenamefont {Levitov}, \citenamefont {Tian}, \citenamefont {{van der Wal CH}},\ and\ \citenamefont {Lloyd}}]{Mooij1999}%
  \BibitemOpen
  \bibfield  {author} {\bibinfo {author} {\bibfnamefont {J.~E.}\ \bibnamefont {Mooij}}, \bibinfo {author} {\bibfnamefont {T.~P.}\ \bibnamefont {Orlando}}, \bibinfo {author} {\bibfnamefont {L.}~\bibnamefont {Levitov}}, \bibinfo {author} {\bibfnamefont {L.}~\bibnamefont {Tian}}, \bibinfo {author} {\bibnamefont {{van der Wal CH}}},\ and\ \bibinfo {author} {\bibfnamefont {S.}~\bibnamefont {Lloyd}},\ }\href {https://doi.org/https://doi.org/10.1126/science.285.5430.1036} {\bibfield  {journal} {\bibinfo  {journal} {Science}\ }\textbf {\bibinfo {volume} {285}},\ \bibinfo {pages} {1036} (\bibinfo {year} {1999})}\BibitemShut {NoStop}%
\bibitem [{\citenamefont {Koch}\ \emph {et~al.}(2007)\citenamefont {Koch}, \citenamefont {Yu}, \citenamefont {Gambetta}, \citenamefont {Houck}, \citenamefont {Schuster}, \citenamefont {Majer}, \citenamefont {Blais}, \citenamefont {Devoret}, \citenamefont {Girvin},\ and\ \citenamefont {Schoelkopf}}]{Koch2007}%
  \BibitemOpen
  \bibfield  {author} {\bibinfo {author} {\bibfnamefont {J.}~\bibnamefont {Koch}}, \bibinfo {author} {\bibfnamefont {T.~M.}\ \bibnamefont {Yu}}, \bibinfo {author} {\bibfnamefont {J.}~\bibnamefont {Gambetta}}, \bibinfo {author} {\bibfnamefont {A.~A.}\ \bibnamefont {Houck}}, \bibinfo {author} {\bibfnamefont {D.~I.}\ \bibnamefont {Schuster}}, \bibinfo {author} {\bibfnamefont {J.}~\bibnamefont {Majer}}, \bibinfo {author} {\bibfnamefont {A.}~\bibnamefont {Blais}}, \bibinfo {author} {\bibfnamefont {M.~H.}\ \bibnamefont {Devoret}}, \bibinfo {author} {\bibfnamefont {S.~M.}\ \bibnamefont {Girvin}},\ and\ \bibinfo {author} {\bibfnamefont {R.~J.}\ \bibnamefont {Schoelkopf}},\ }\href {https://doi.org/10.1103/PhysRevA.76.042319} {\bibfield  {journal} {\bibinfo  {journal} {Phys. Rev. A}\ }\textbf {\bibinfo {volume} {76}},\ \bibinfo {pages} {042319} (\bibinfo {year} {2007})}\BibitemShut {NoStop}%
\bibitem [{\citenamefont {Yan}\ \emph {et~al.}(2016)\citenamefont {Yan}, \citenamefont {Gustavsson}, \citenamefont {Kamal}, \citenamefont {Birenbaum}, \citenamefont {Sears}, \citenamefont {Hover}, \citenamefont {Gudmundsen}, \citenamefont {Rosenberg}, \citenamefont {Samach}, \citenamefont {Weber}, \citenamefont {Yoder}, \citenamefont {Orlando}, \citenamefont {Clarke}, \citenamefont {Kerman},\ and\ \citenamefont {Oliver}}]{Yan2016}%
  \BibitemOpen
  \bibfield  {author} {\bibinfo {author} {\bibfnamefont {F.}~\bibnamefont {Yan}}, \bibinfo {author} {\bibfnamefont {S.}~\bibnamefont {Gustavsson}}, \bibinfo {author} {\bibfnamefont {A.}~\bibnamefont {Kamal}}, \bibinfo {author} {\bibfnamefont {J.}~\bibnamefont {Birenbaum}}, \bibinfo {author} {\bibfnamefont {A.~P.}\ \bibnamefont {Sears}}, \bibinfo {author} {\bibfnamefont {D.}~\bibnamefont {Hover}}, \bibinfo {author} {\bibfnamefont {T.~J.}\ \bibnamefont {Gudmundsen}}, \bibinfo {author} {\bibfnamefont {D.}~\bibnamefont {Rosenberg}}, \bibinfo {author} {\bibfnamefont {G.}~\bibnamefont {Samach}}, \bibinfo {author} {\bibfnamefont {S.}~\bibnamefont {Weber}}, \bibinfo {author} {\bibfnamefont {J.~L.}\ \bibnamefont {Yoder}}, \bibinfo {author} {\bibfnamefont {T.~P.}\ \bibnamefont {Orlando}}, \bibinfo {author} {\bibfnamefont {J.}~\bibnamefont {Clarke}}, \bibinfo {author} {\bibfnamefont {A.~J.}\ \bibnamefont {Kerman}},\ and\ \bibinfo {author} {\bibfnamefont {W.~D.}\ \bibnamefont {Oliver}},\ }\href
  {https://doi.org/https://doi.org/10.1038/ncomms12964} {\bibfield  {journal} {\bibinfo  {journal} {Nat. Commun.}\ }\textbf {\bibinfo {volume} {7}},\ \bibinfo {pages} {12964} (\bibinfo {year} {2016})}\BibitemShut {NoStop}%
\bibitem [{\citenamefont {Gyenis}\ \emph {et~al.}(2021)\citenamefont {Gyenis}, \citenamefont {Di~Paolo}, \citenamefont {Koch}, \citenamefont {Blais}, \citenamefont {Houck},\ and\ \citenamefont {Schuster}}]{Gyenis2021}%
  \BibitemOpen
  \bibfield  {author} {\bibinfo {author} {\bibfnamefont {A.}~\bibnamefont {Gyenis}}, \bibinfo {author} {\bibfnamefont {A.}~\bibnamefont {Di~Paolo}}, \bibinfo {author} {\bibfnamefont {J.}~\bibnamefont {Koch}}, \bibinfo {author} {\bibfnamefont {A.}~\bibnamefont {Blais}}, \bibinfo {author} {\bibfnamefont {A.~A.}\ \bibnamefont {Houck}},\ and\ \bibinfo {author} {\bibfnamefont {D.~I.}\ \bibnamefont {Schuster}},\ }\href {https://doi.org/10.1103/PRXQuantum.2.030101} {\bibfield  {journal} {\bibinfo  {journal} {PRX Quantum}\ }\textbf {\bibinfo {volume} {2}},\ \bibinfo {pages} {030101} (\bibinfo {year} {2021})}\BibitemShut {NoStop}%
\bibitem [{\citenamefont {Arute}\ \emph {et~al.}(2019)\citenamefont {Arute}, \citenamefont {Arya}, \citenamefont {Babbush}, \citenamefont {Bacon}, \citenamefont {Bardin}, \citenamefont {Barends}, \citenamefont {Biswas}, \citenamefont {Boixo}, \citenamefont {Brandao}, \citenamefont {Buell}, \citenamefont {Burkett}, \citenamefont {Chen}, \citenamefont {Chen}, \citenamefont {Chiaro}, \citenamefont {Collins}, \citenamefont {Courtney}, \citenamefont {Dunsworth}, \citenamefont {Farhi}, \citenamefont {Foxen}, \citenamefont {Fowler}, \citenamefont {Gidney}, \citenamefont {Giustina}, \citenamefont {Graff}, \citenamefont {Guerin}, \citenamefont {Habegger}, \citenamefont {Harrigan}, \citenamefont {Hartmann}, \citenamefont {Ho}, \citenamefont {Hoffmann}, \citenamefont {Huang}, \citenamefont {Humble}, \citenamefont {Isakov}, \citenamefont {Jeffrey}, \citenamefont {Jiang}, \citenamefont {Kafri}, \citenamefont {Kechedzhi}, \citenamefont {Kelly}, \citenamefont {Klimov}, \citenamefont {Knysh}, \citenamefont {Korotkov},
  \citenamefont {Kostritsa}, \citenamefont {Landhuis}, \citenamefont {Lindmark}, \citenamefont {Lucero}, \citenamefont {Lyakh}, \citenamefont {Mandr{\`{a}}}, \citenamefont {McClean}, \citenamefont {McEwen}, \citenamefont {Megrant}, \citenamefont {Mi}, \citenamefont {Michielsen}, \citenamefont {Mohseni}, \citenamefont {Mutus}, \citenamefont {Naaman}, \citenamefont {Neeley}, \citenamefont {Neill}, \citenamefont {Niu}, \citenamefont {Ostby}, \citenamefont {Petukhov}, \citenamefont {Platt}, \citenamefont {Quintana}, \citenamefont {Rieffel}, \citenamefont {Roushan}, \citenamefont {Rubin}, \citenamefont {Sank}, \citenamefont {Satzinger}, \citenamefont {Smelyanskiy}, \citenamefont {Sung}, \citenamefont {Trevithick}, \citenamefont {Vainsencher}, \citenamefont {Villalonga}, \citenamefont {White}, \citenamefont {Yao}, \citenamefont {Yeh}, \citenamefont {Zalcman}, \citenamefont {Neven},\ and\ \citenamefont {Martinis}}]{Arute2019}%
  \BibitemOpen
  \bibfield  {author} {\bibinfo {author} {\bibfnamefont {F.}~\bibnamefont {Arute}}, \bibinfo {author} {\bibfnamefont {K.}~\bibnamefont {Arya}}, \bibinfo {author} {\bibfnamefont {R.}~\bibnamefont {Babbush}}, \bibinfo {author} {\bibfnamefont {D.}~\bibnamefont {Bacon}}, \bibinfo {author} {\bibfnamefont {J.~C.}\ \bibnamefont {Bardin}}, \bibinfo {author} {\bibfnamefont {R.}~\bibnamefont {Barends}}, \bibinfo {author} {\bibfnamefont {R.}~\bibnamefont {Biswas}}, \bibinfo {author} {\bibfnamefont {S.}~\bibnamefont {Boixo}}, \bibinfo {author} {\bibfnamefont {F.~G. S.~L.}\ \bibnamefont {Brandao}}, \bibinfo {author} {\bibfnamefont {D.~A.}\ \bibnamefont {Buell}}, \bibinfo {author} {\bibfnamefont {B.}~\bibnamefont {Burkett}}, \bibinfo {author} {\bibfnamefont {Y.}~\bibnamefont {Chen}}, \bibinfo {author} {\bibfnamefont {Z.}~\bibnamefont {Chen}}, \bibinfo {author} {\bibfnamefont {B.}~\bibnamefont {Chiaro}}, \bibinfo {author} {\bibfnamefont {R.}~\bibnamefont {Collins}}, \bibinfo {author} {\bibfnamefont {W.}~\bibnamefont
  {Courtney}}, \bibinfo {author} {\bibfnamefont {A.}~\bibnamefont {Dunsworth}}, \bibinfo {author} {\bibfnamefont {E.}~\bibnamefont {Farhi}}, \bibinfo {author} {\bibfnamefont {B.}~\bibnamefont {Foxen}}, \bibinfo {author} {\bibfnamefont {A.}~\bibnamefont {Fowler}}, \bibinfo {author} {\bibfnamefont {C.}~\bibnamefont {Gidney}}, \bibinfo {author} {\bibfnamefont {M.}~\bibnamefont {Giustina}}, \bibinfo {author} {\bibfnamefont {R.}~\bibnamefont {Graff}}, \bibinfo {author} {\bibfnamefont {K.}~\bibnamefont {Guerin}}, \bibinfo {author} {\bibfnamefont {S.}~\bibnamefont {Habegger}}, \bibinfo {author} {\bibfnamefont {M.~P.}\ \bibnamefont {Harrigan}}, \bibinfo {author} {\bibfnamefont {M.~J.}\ \bibnamefont {Hartmann}}, \bibinfo {author} {\bibfnamefont {A.}~\bibnamefont {Ho}}, \bibinfo {author} {\bibfnamefont {M.}~\bibnamefont {Hoffmann}}, \bibinfo {author} {\bibfnamefont {T.}~\bibnamefont {Huang}}, \bibinfo {author} {\bibfnamefont {T.~S.}\ \bibnamefont {Humble}}, \bibinfo {author} {\bibfnamefont {S.~V.}\ \bibnamefont
  {Isakov}}, \bibinfo {author} {\bibfnamefont {E.}~\bibnamefont {Jeffrey}}, \bibinfo {author} {\bibfnamefont {Z.}~\bibnamefont {Jiang}}, \bibinfo {author} {\bibfnamefont {D.}~\bibnamefont {Kafri}}, \bibinfo {author} {\bibfnamefont {K.}~\bibnamefont {Kechedzhi}}, \bibinfo {author} {\bibfnamefont {J.}~\bibnamefont {Kelly}}, \bibinfo {author} {\bibfnamefont {P.~V.}\ \bibnamefont {Klimov}}, \bibinfo {author} {\bibfnamefont {S.}~\bibnamefont {Knysh}}, \bibinfo {author} {\bibfnamefont {A.}~\bibnamefont {Korotkov}}, \bibinfo {author} {\bibfnamefont {F.}~\bibnamefont {Kostritsa}}, \bibinfo {author} {\bibfnamefont {D.}~\bibnamefont {Landhuis}}, \bibinfo {author} {\bibfnamefont {M.}~\bibnamefont {Lindmark}}, \bibinfo {author} {\bibfnamefont {E.}~\bibnamefont {Lucero}}, \bibinfo {author} {\bibfnamefont {D.}~\bibnamefont {Lyakh}}, \bibinfo {author} {\bibfnamefont {S.}~\bibnamefont {Mandr{\`{a}}}}, \bibinfo {author} {\bibfnamefont {J.~R.}\ \bibnamefont {McClean}}, \bibinfo {author} {\bibfnamefont {M.}~\bibnamefont
  {McEwen}}, \bibinfo {author} {\bibfnamefont {A.}~\bibnamefont {Megrant}}, \bibinfo {author} {\bibfnamefont {X.}~\bibnamefont {Mi}}, \bibinfo {author} {\bibfnamefont {K.}~\bibnamefont {Michielsen}}, \bibinfo {author} {\bibfnamefont {M.}~\bibnamefont {Mohseni}}, \bibinfo {author} {\bibfnamefont {J.}~\bibnamefont {Mutus}}, \bibinfo {author} {\bibfnamefont {O.}~\bibnamefont {Naaman}}, \bibinfo {author} {\bibfnamefont {M.}~\bibnamefont {Neeley}}, \bibinfo {author} {\bibfnamefont {C.}~\bibnamefont {Neill}}, \bibinfo {author} {\bibfnamefont {M.~Y.}\ \bibnamefont {Niu}}, \bibinfo {author} {\bibfnamefont {E.}~\bibnamefont {Ostby}}, \bibinfo {author} {\bibfnamefont {A.}~\bibnamefont {Petukhov}}, \bibinfo {author} {\bibfnamefont {J.~C.}\ \bibnamefont {Platt}}, \bibinfo {author} {\bibfnamefont {C.}~\bibnamefont {Quintana}}, \bibinfo {author} {\bibfnamefont {E.~G.}\ \bibnamefont {Rieffel}}, \bibinfo {author} {\bibfnamefont {P.}~\bibnamefont {Roushan}}, \bibinfo {author} {\bibfnamefont {N.~C.}\ \bibnamefont {Rubin}},
  \bibinfo {author} {\bibfnamefont {D.}~\bibnamefont {Sank}}, \bibinfo {author} {\bibfnamefont {K.~J.}\ \bibnamefont {Satzinger}}, \bibinfo {author} {\bibfnamefont {V.}~\bibnamefont {Smelyanskiy}}, \bibinfo {author} {\bibfnamefont {K.~J.}\ \bibnamefont {Sung}}, \bibinfo {author} {\bibfnamefont {M.~D.}\ \bibnamefont {Trevithick}}, \bibinfo {author} {\bibfnamefont {A.}~\bibnamefont {Vainsencher}}, \bibinfo {author} {\bibfnamefont {B.}~\bibnamefont {Villalonga}}, \bibinfo {author} {\bibfnamefont {T.}~\bibnamefont {White}}, \bibinfo {author} {\bibfnamefont {Z.~J.}\ \bibnamefont {Yao}}, \bibinfo {author} {\bibfnamefont {P.}~\bibnamefont {Yeh}}, \bibinfo {author} {\bibfnamefont {A.}~\bibnamefont {Zalcman}}, \bibinfo {author} {\bibfnamefont {H.}~\bibnamefont {Neven}},\ and\ \bibinfo {author} {\bibfnamefont {J.~M.}\ \bibnamefont {Martinis}},\ }\href {https://doi.org/10.1038/s41586-019-1666-5} {\bibfield  {journal} {\bibinfo  {journal} {Nature}\ }\textbf {\bibinfo {volume} {574}},\ \bibinfo {pages} {505} (\bibinfo
  {year} {2019})}\BibitemShut {NoStop}%
\bibitem [{\citenamefont {Jurcevic}\ \emph {et~al.}(2021)\citenamefont {Jurcevic}, \citenamefont {Javadi-Abhari}, \citenamefont {Bishop}, \citenamefont {Lauer}, \citenamefont {Bogorin}, \citenamefont {Brink}, \citenamefont {Capelluto}, \citenamefont {G{\"{u}}nl{\"{u}}k}, \citenamefont {Itoko}, \citenamefont {Kanazawa}, \citenamefont {Kandala}, \citenamefont {Keefe}, \citenamefont {Krsulich}, \citenamefont {Landers}, \citenamefont {Lewandowski}, \citenamefont {McClure}, \citenamefont {Nannicini}, \citenamefont {Narasgond}, \citenamefont {Nayfeh}, \citenamefont {Pritchett}, \citenamefont {Rothwell}, \citenamefont {Srinivasan}, \citenamefont {Sundaresan}, \citenamefont {Wang}, \citenamefont {Wei}, \citenamefont {Wood}, \citenamefont {Yau}, \citenamefont {Zhang}, \citenamefont {Dial}, \citenamefont {Chow},\ and\ \citenamefont {Gambetta}}]{Jurcevic2021}%
  \BibitemOpen
  \bibfield  {author} {\bibinfo {author} {\bibfnamefont {P.}~\bibnamefont {Jurcevic}}, \bibinfo {author} {\bibfnamefont {A.}~\bibnamefont {Javadi-Abhari}}, \bibinfo {author} {\bibfnamefont {L.~S.}\ \bibnamefont {Bishop}}, \bibinfo {author} {\bibfnamefont {I.}~\bibnamefont {Lauer}}, \bibinfo {author} {\bibfnamefont {D.~F.}\ \bibnamefont {Bogorin}}, \bibinfo {author} {\bibfnamefont {M.}~\bibnamefont {Brink}}, \bibinfo {author} {\bibfnamefont {L.}~\bibnamefont {Capelluto}}, \bibinfo {author} {\bibfnamefont {O.}~\bibnamefont {G{\"{u}}nl{\"{u}}k}}, \bibinfo {author} {\bibfnamefont {T.}~\bibnamefont {Itoko}}, \bibinfo {author} {\bibfnamefont {N.}~\bibnamefont {Kanazawa}}, \bibinfo {author} {\bibfnamefont {A.}~\bibnamefont {Kandala}}, \bibinfo {author} {\bibfnamefont {G.~A.}\ \bibnamefont {Keefe}}, \bibinfo {author} {\bibfnamefont {K.}~\bibnamefont {Krsulich}}, \bibinfo {author} {\bibfnamefont {W.}~\bibnamefont {Landers}}, \bibinfo {author} {\bibfnamefont {E.~P.}\ \bibnamefont {Lewandowski}}, \bibinfo {author}
  {\bibfnamefont {D.~T.}\ \bibnamefont {McClure}}, \bibinfo {author} {\bibfnamefont {G.}~\bibnamefont {Nannicini}}, \bibinfo {author} {\bibfnamefont {A.}~\bibnamefont {Narasgond}}, \bibinfo {author} {\bibfnamefont {H.~M.}\ \bibnamefont {Nayfeh}}, \bibinfo {author} {\bibfnamefont {E.}~\bibnamefont {Pritchett}}, \bibinfo {author} {\bibfnamefont {M.~B.}\ \bibnamefont {Rothwell}}, \bibinfo {author} {\bibfnamefont {S.}~\bibnamefont {Srinivasan}}, \bibinfo {author} {\bibfnamefont {N.}~\bibnamefont {Sundaresan}}, \bibinfo {author} {\bibfnamefont {C.}~\bibnamefont {Wang}}, \bibinfo {author} {\bibfnamefont {K.~X.}\ \bibnamefont {Wei}}, \bibinfo {author} {\bibfnamefont {C.~J.}\ \bibnamefont {Wood}}, \bibinfo {author} {\bibfnamefont {J.-B.}\ \bibnamefont {Yau}}, \bibinfo {author} {\bibfnamefont {E.~J.}\ \bibnamefont {Zhang}}, \bibinfo {author} {\bibfnamefont {O.~E.}\ \bibnamefont {Dial}}, \bibinfo {author} {\bibfnamefont {J.~M.}\ \bibnamefont {Chow}},\ and\ \bibinfo {author} {\bibfnamefont {J.~M.}\ \bibnamefont
  {Gambetta}},\ }\href {https://doi.org/10.1088/2058-9565/abe519} {\bibfield  {journal} {\bibinfo  {journal} {Quantum Science and Technology}\ }\textbf {\bibinfo {volume} {6}},\ \bibinfo {pages} {025020} (\bibinfo {year} {2021})}\BibitemShut {NoStop}%
\bibitem [{\citenamefont {{Google Quantum AI and Collaborators}}(2025)}]{Acharya2024}%
  \BibitemOpen
  \bibfield  {author} {\bibinfo {author} {\bibnamefont {{Google Quantum AI and Collaborators}}},\ }\href {https://doi.org/https://doi.org/10.1038/s41586-024-08449-y} {\bibfield  {journal} {\bibinfo  {journal} {Nature}\ }\textbf {\bibinfo {volume} {638}},\ \bibinfo {pages} {920} (\bibinfo {year} {2025})}\BibitemShut {NoStop}%
\bibitem [{\citenamefont {Manucharyan}\ \emph {et~al.}(2009)\citenamefont {Manucharyan}, \citenamefont {Koch}, \citenamefont {Glazman},\ and\ \citenamefont {Devoret}}]{Manucharyan2009}%
  \BibitemOpen
  \bibfield  {author} {\bibinfo {author} {\bibfnamefont {V.~E.}\ \bibnamefont {Manucharyan}}, \bibinfo {author} {\bibfnamefont {J.}~\bibnamefont {Koch}}, \bibinfo {author} {\bibfnamefont {L.~I.}\ \bibnamefont {Glazman}},\ and\ \bibinfo {author} {\bibfnamefont {M.~H.}\ \bibnamefont {Devoret}},\ }\href {https://doi.org/https://www.science.org/doi/10.1126/science.1175552} {\bibfield  {journal} {\bibinfo  {journal} {Science}\ }\textbf {\bibinfo {volume} {326}},\ \bibinfo {pages} {113} (\bibinfo {year} {2009})}\BibitemShut {NoStop}%
\bibitem [{\citenamefont {Nguyen}\ \emph {et~al.}(2022)\citenamefont {Nguyen}, \citenamefont {Koolstra}, \citenamefont {Kim}, \citenamefont {Morvan}, \citenamefont {Chistolini}, \citenamefont {Singh}, \citenamefont {Nesterov}, \citenamefont {J\"unger}, \citenamefont {Chen}, \citenamefont {Pedramrazi}, \citenamefont {Mitchell}, \citenamefont {Kreikebaum}, \citenamefont {Puri}, \citenamefont {Santiago},\ and\ \citenamefont {Siddiqi}}]{Nguyen2022}%
  \BibitemOpen
  \bibfield  {author} {\bibinfo {author} {\bibfnamefont {L.~B.}\ \bibnamefont {Nguyen}}, \bibinfo {author} {\bibfnamefont {G.}~\bibnamefont {Koolstra}}, \bibinfo {author} {\bibfnamefont {Y.}~\bibnamefont {Kim}}, \bibinfo {author} {\bibfnamefont {A.}~\bibnamefont {Morvan}}, \bibinfo {author} {\bibfnamefont {T.}~\bibnamefont {Chistolini}}, \bibinfo {author} {\bibfnamefont {S.}~\bibnamefont {Singh}}, \bibinfo {author} {\bibfnamefont {K.~N.}\ \bibnamefont {Nesterov}}, \bibinfo {author} {\bibfnamefont {C.}~\bibnamefont {J\"unger}}, \bibinfo {author} {\bibfnamefont {L.}~\bibnamefont {Chen}}, \bibinfo {author} {\bibfnamefont {Z.}~\bibnamefont {Pedramrazi}}, \bibinfo {author} {\bibfnamefont {B.~K.}\ \bibnamefont {Mitchell}}, \bibinfo {author} {\bibfnamefont {J.~M.}\ \bibnamefont {Kreikebaum}}, \bibinfo {author} {\bibfnamefont {S.}~\bibnamefont {Puri}}, \bibinfo {author} {\bibfnamefont {D.~I.}\ \bibnamefont {Santiago}},\ and\ \bibinfo {author} {\bibfnamefont {I.}~\bibnamefont {Siddiqi}},\ }\href
  {https://doi.org/10.1103/PRXQuantum.3.037001} {\bibfield  {journal} {\bibinfo  {journal} {PRX Quantum}\ }\textbf {\bibinfo {volume} {3}},\ \bibinfo {pages} {037001} (\bibinfo {year} {2022})}\BibitemShut {NoStop}%
\bibitem [{\citenamefont {Somoroff}\ \emph {et~al.}(2023)\citenamefont {Somoroff}, \citenamefont {Ficheux}, \citenamefont {Mencia}, \citenamefont {Xiong}, \citenamefont {Kuzmin},\ and\ \citenamefont {Manucharyan}}]{Somoroff2023}%
  \BibitemOpen
  \bibfield  {author} {\bibinfo {author} {\bibfnamefont {A.}~\bibnamefont {Somoroff}}, \bibinfo {author} {\bibfnamefont {Q.}~\bibnamefont {Ficheux}}, \bibinfo {author} {\bibfnamefont {R.~A.}\ \bibnamefont {Mencia}}, \bibinfo {author} {\bibfnamefont {H.}~\bibnamefont {Xiong}}, \bibinfo {author} {\bibfnamefont {R.}~\bibnamefont {Kuzmin}},\ and\ \bibinfo {author} {\bibfnamefont {V.~E.}\ \bibnamefont {Manucharyan}},\ }\href {https://doi.org/https://doi.org/10.1103/PhysRevLett.130.267001} {\bibfield  {journal} {\bibinfo  {journal} {Phys. Rev. Lett.}\ }\textbf {\bibinfo {volume} {130}},\ \bibinfo {pages} {267001} (\bibinfo {year} {2023})}\BibitemShut {NoStop}%
\bibitem [{\citenamefont {Wang}\ \emph {et~al.}(2024{\natexlab{a}})\citenamefont {Wang}, \citenamefont {Lu}, \citenamefont {Zhan}, \citenamefont {Ma}, \citenamefont {Wu}, \citenamefont {Sun}, \citenamefont {Deng}, \citenamefont {Bai}, \citenamefont {Bao}, \citenamefont {Chang}, \citenamefont {Gao}, \citenamefont {Gao}, \citenamefont {Gong}, \citenamefont {Hu}, \citenamefont {Hu}, \citenamefont {Ji}, \citenamefont {Ma}, \citenamefont {Mao}, \citenamefont {Song}, \citenamefont {Tang}, \citenamefont {Wang}, \citenamefont {Wang}, \citenamefont {Wang}, \citenamefont {Xia}, \citenamefont {Xu}, \citenamefont {Zhan}, \citenamefont {Zhang}, \citenamefont {Zhou}, \citenamefont {Zhu}, \citenamefont {Zhu}, \citenamefont {Zhu}, \citenamefont {Zhu}, \citenamefont {Shi}, \citenamefont {Zhao},\ and\ \citenamefont {Deng}}]{FeiWang2024}%
  \BibitemOpen
  \bibfield  {author} {\bibinfo {author} {\bibfnamefont {F.}~\bibnamefont {Wang}}, \bibinfo {author} {\bibfnamefont {K.}~\bibnamefont {Lu}}, \bibinfo {author} {\bibfnamefont {H.}~\bibnamefont {Zhan}}, \bibinfo {author} {\bibfnamefont {L.}~\bibnamefont {Ma}}, \bibinfo {author} {\bibfnamefont {F.}~\bibnamefont {Wu}}, \bibinfo {author} {\bibfnamefont {H.}~\bibnamefont {Sun}}, \bibinfo {author} {\bibfnamefont {H.}~\bibnamefont {Deng}}, \bibinfo {author} {\bibfnamefont {Y.}~\bibnamefont {Bai}}, \bibinfo {author} {\bibfnamefont {F.}~\bibnamefont {Bao}}, \bibinfo {author} {\bibfnamefont {X.}~\bibnamefont {Chang}}, \bibinfo {author} {\bibfnamefont {R.}~\bibnamefont {Gao}}, \bibinfo {author} {\bibfnamefont {X.}~\bibnamefont {Gao}}, \bibinfo {author} {\bibfnamefont {G.}~\bibnamefont {Gong}}, \bibinfo {author} {\bibfnamefont {L.}~\bibnamefont {Hu}}, \bibinfo {author} {\bibfnamefont {R.}~\bibnamefont {Hu}}, \bibinfo {author} {\bibfnamefont {H.}~\bibnamefont {Ji}}, \bibinfo {author} {\bibfnamefont {X.}~\bibnamefont {Ma}},
  \bibinfo {author} {\bibfnamefont {L.}~\bibnamefont {Mao}}, \bibinfo {author} {\bibfnamefont {Z.}~\bibnamefont {Song}}, \bibinfo {author} {\bibfnamefont {C.}~\bibnamefont {Tang}}, \bibinfo {author} {\bibfnamefont {H.}~\bibnamefont {Wang}}, \bibinfo {author} {\bibfnamefont {T.}~\bibnamefont {Wang}}, \bibinfo {author} {\bibfnamefont {Z.}~\bibnamefont {Wang}}, \bibinfo {author} {\bibfnamefont {T.}~\bibnamefont {Xia}}, \bibinfo {author} {\bibfnamefont {H.}~\bibnamefont {Xu}}, \bibinfo {author} {\bibfnamefont {Z.}~\bibnamefont {Zhan}}, \bibinfo {author} {\bibfnamefont {G.}~\bibnamefont {Zhang}}, \bibinfo {author} {\bibfnamefont {T.}~\bibnamefont {Zhou}}, \bibinfo {author} {\bibfnamefont {M.}~\bibnamefont {Zhu}}, \bibinfo {author} {\bibfnamefont {Q.}~\bibnamefont {Zhu}}, \bibinfo {author} {\bibfnamefont {S.}~\bibnamefont {Zhu}}, \bibinfo {author} {\bibfnamefont {X.}~\bibnamefont {Zhu}}, \bibinfo {author} {\bibfnamefont {Y.}~\bibnamefont {Shi}}, \bibinfo {author} {\bibfnamefont {H.-H.}\ \bibnamefont {Zhao}},\ and\
  \bibinfo {author} {\bibfnamefont {C.}~\bibnamefont {Deng}},\ }\href@noop {} {\bibinfo {title} {Achieving millisecond coherence fluxonium through overlap {Josephson} junctions}} (\bibinfo {year} {2024}{\natexlab{a}}),\ \Eprint {https://arxiv.org/abs/2405.05481} {arXiv:2405.05481 [quant-ph]} \BibitemShut {NoStop}%
\bibitem [{\citenamefont {Rower}\ \emph {et~al.}(2024)\citenamefont {Rower}, \citenamefont {Ding}, \citenamefont {Zhang}, \citenamefont {Hays}, \citenamefont {An}, \citenamefont {Harrington}, \citenamefont {Rosen}, \citenamefont {Gertler}, \citenamefont {Hazard}, \citenamefont {Niedzielski}, \citenamefont {Schwartz}, \citenamefont {Gustavsson}, \citenamefont {Serniak}, \citenamefont {Grover},\ and\ \citenamefont {Oliver}}]{Rower2024}%
  \BibitemOpen
  \bibfield  {author} {\bibinfo {author} {\bibfnamefont {D.~A.}\ \bibnamefont {Rower}}, \bibinfo {author} {\bibfnamefont {L.}~\bibnamefont {Ding}}, \bibinfo {author} {\bibfnamefont {H.}~\bibnamefont {Zhang}}, \bibinfo {author} {\bibfnamefont {M.}~\bibnamefont {Hays}}, \bibinfo {author} {\bibfnamefont {J.}~\bibnamefont {An}}, \bibinfo {author} {\bibfnamefont {P.~M.}\ \bibnamefont {Harrington}}, \bibinfo {author} {\bibfnamefont {I.~T.}\ \bibnamefont {Rosen}}, \bibinfo {author} {\bibfnamefont {J.~M.}\ \bibnamefont {Gertler}}, \bibinfo {author} {\bibfnamefont {T.~M.}\ \bibnamefont {Hazard}}, \bibinfo {author} {\bibfnamefont {B.~M.}\ \bibnamefont {Niedzielski}}, \bibinfo {author} {\bibfnamefont {M.~E.}\ \bibnamefont {Schwartz}}, \bibinfo {author} {\bibfnamefont {S.}~\bibnamefont {Gustavsson}}, \bibinfo {author} {\bibfnamefont {K.}~\bibnamefont {Serniak}}, \bibinfo {author} {\bibfnamefont {J.~A.}\ \bibnamefont {Grover}},\ and\ \bibinfo {author} {\bibfnamefont {W.~D.}\ \bibnamefont {Oliver}},\ }\href
  {https://doi.org/10.1103/PRXQuantum.5.040342} {\bibfield  {journal} {\bibinfo  {journal} {PRX Quantum}\ }\textbf {\bibinfo {volume} {5}},\ \bibinfo {pages} {040342} (\bibinfo {year} {2024})}\BibitemShut {NoStop}%
\bibitem [{\citenamefont {Lin}\ \emph {et~al.}(2025{\natexlab{a}})\citenamefont {Lin}, \citenamefont {Cho}, \citenamefont {Chen}, \citenamefont {Vavilov}, \citenamefont {Wang},\ and\ \citenamefont {Manucharyan}}]{Lin2024}%
  \BibitemOpen
  \bibfield  {author} {\bibinfo {author} {\bibfnamefont {W.-J.}\ \bibnamefont {Lin}}, \bibinfo {author} {\bibfnamefont {H.}~\bibnamefont {Cho}}, \bibinfo {author} {\bibfnamefont {Y.}~\bibnamefont {Chen}}, \bibinfo {author} {\bibfnamefont {M.~G.}\ \bibnamefont {Vavilov}}, \bibinfo {author} {\bibfnamefont {C.}~\bibnamefont {Wang}},\ and\ \bibinfo {author} {\bibfnamefont {V.~E.}\ \bibnamefont {Manucharyan}},\ }\href {https://doi.org/10.1103/PRXQuantum.6.010349} {\bibfield  {journal} {\bibinfo  {journal} {PRX Quantum}\ }\textbf {\bibinfo {volume} {6}},\ \bibinfo {pages} {010349} (\bibinfo {year} {2025}{\natexlab{a}})}\BibitemShut {NoStop}%
\bibitem [{\citenamefont {Stefanski}\ and\ \citenamefont {Andersen}(2024)}]{Stefanski2024}%
  \BibitemOpen
  \bibfield  {author} {\bibinfo {author} {\bibfnamefont {T.~V.}\ \bibnamefont {Stefanski}}\ and\ \bibinfo {author} {\bibfnamefont {C.~K.}\ \bibnamefont {Andersen}},\ }\href {https://doi.org/10.1103/PhysRevApplied.22.014079} {\bibfield  {journal} {\bibinfo  {journal} {Phys. Rev. Appl.}\ }\textbf {\bibinfo {volume} {22}},\ \bibinfo {pages} {014079} (\bibinfo {year} {2024})}\BibitemShut {NoStop}%
\bibitem [{\citenamefont {Schirk}\ \emph {et~al.}(2024)\citenamefont {Schirk}, \citenamefont {Wallner}, \citenamefont {Huang}, \citenamefont {Tsitsilin}, \citenamefont {Bruckmoser}, \citenamefont {Koch}, \citenamefont {Bunch}, \citenamefont {Glaser}, \citenamefont {Huber}, \citenamefont {Knudsen}, \citenamefont {Krylov}, \citenamefont {Marx}, \citenamefont {Pfeiffer}, \citenamefont {Richard}, \citenamefont {Roy}, \citenamefont {Romeiro}, \citenamefont {Singh}, \citenamefont {Södergren}, \citenamefont {Dionis}, \citenamefont {Sugny}, \citenamefont {Werninghaus}, \citenamefont {Liegener}, \citenamefont {Schneider},\ and\ \citenamefont {Filipp}}]{Schirk2024}%
  \BibitemOpen
  \bibfield  {author} {\bibinfo {author} {\bibfnamefont {J.}~\bibnamefont {Schirk}}, \bibinfo {author} {\bibfnamefont {F.}~\bibnamefont {Wallner}}, \bibinfo {author} {\bibfnamefont {L.}~\bibnamefont {Huang}}, \bibinfo {author} {\bibfnamefont {I.}~\bibnamefont {Tsitsilin}}, \bibinfo {author} {\bibfnamefont {N.}~\bibnamefont {Bruckmoser}}, \bibinfo {author} {\bibfnamefont {L.}~\bibnamefont {Koch}}, \bibinfo {author} {\bibfnamefont {D.}~\bibnamefont {Bunch}}, \bibinfo {author} {\bibfnamefont {N.~J.}\ \bibnamefont {Glaser}}, \bibinfo {author} {\bibfnamefont {G.~B.~P.}\ \bibnamefont {Huber}}, \bibinfo {author} {\bibfnamefont {M.}~\bibnamefont {Knudsen}}, \bibinfo {author} {\bibfnamefont {G.}~\bibnamefont {Krylov}}, \bibinfo {author} {\bibfnamefont {A.}~\bibnamefont {Marx}}, \bibinfo {author} {\bibfnamefont {F.}~\bibnamefont {Pfeiffer}}, \bibinfo {author} {\bibfnamefont {L.}~\bibnamefont {Richard}}, \bibinfo {author} {\bibfnamefont {F.~A.}\ \bibnamefont {Roy}}, \bibinfo {author} {\bibfnamefont {J.~H.}\ \bibnamefont
  {Romeiro}}, \bibinfo {author} {\bibfnamefont {M.}~\bibnamefont {Singh}}, \bibinfo {author} {\bibfnamefont {L.}~\bibnamefont {Södergren}}, \bibinfo {author} {\bibfnamefont {E.}~\bibnamefont {Dionis}}, \bibinfo {author} {\bibfnamefont {D.}~\bibnamefont {Sugny}}, \bibinfo {author} {\bibfnamefont {M.}~\bibnamefont {Werninghaus}}, \bibinfo {author} {\bibfnamefont {K.}~\bibnamefont {Liegener}}, \bibinfo {author} {\bibfnamefont {C.~M.~F.}\ \bibnamefont {Schneider}},\ and\ \bibinfo {author} {\bibfnamefont {S.}~\bibnamefont {Filipp}},\ }\href@noop {} {\bibinfo {title} {Protected fluxonium control with sub-harmonic parametric driving}} (\bibinfo {year} {2024}),\ \Eprint {https://arxiv.org/abs/2410.00495} {arXiv:2410.00495 [quant-ph]} \BibitemShut {NoStop}%
\bibitem [{\citenamefont {Lin}\ \emph {et~al.}(2025{\natexlab{b}})\citenamefont {Lin}, \citenamefont {Cho}, \citenamefont {Chen}, \citenamefont {Vavilov}, \citenamefont {Wang},\ and\ \citenamefont {Manucharyan}}]{Lin2025}%
  \BibitemOpen
  \bibfield  {author} {\bibinfo {author} {\bibfnamefont {W.-J.}\ \bibnamefont {Lin}}, \bibinfo {author} {\bibfnamefont {H.}~\bibnamefont {Cho}}, \bibinfo {author} {\bibfnamefont {Y.}~\bibnamefont {Chen}}, \bibinfo {author} {\bibfnamefont {M.~G.}\ \bibnamefont {Vavilov}}, \bibinfo {author} {\bibfnamefont {C.}~\bibnamefont {Wang}},\ and\ \bibinfo {author} {\bibfnamefont {V.~E.}\ \bibnamefont {Manucharyan}},\ }\href@noop {} {\bibfield  {journal} {\bibinfo  {journal} {New J. Phys.}\ }\textbf {\bibinfo {volume} {27}},\ \bibinfo {pages} {033012} (\bibinfo {year} {2025}{\natexlab{b}})}\BibitemShut {NoStop}%
\bibitem [{\citenamefont {Blais}\ \emph {et~al.}(2021)\citenamefont {Blais}, \citenamefont {Grimsmo}, \citenamefont {Girvin},\ and\ \citenamefont {Wallraff}}]{Blais2021}%
  \BibitemOpen
  \bibfield  {author} {\bibinfo {author} {\bibfnamefont {A.}~\bibnamefont {Blais}}, \bibinfo {author} {\bibfnamefont {A.~L.}\ \bibnamefont {Grimsmo}}, \bibinfo {author} {\bibfnamefont {S.~M.}\ \bibnamefont {Girvin}},\ and\ \bibinfo {author} {\bibfnamefont {A.}~\bibnamefont {Wallraff}},\ }\href {https://doi.org/10.1103/RevModPhys.93.025005} {\bibfield  {journal} {\bibinfo  {journal} {Rev. Mod. Phys.}\ }\textbf {\bibinfo {volume} {93}},\ \bibinfo {pages} {025005} (\bibinfo {year} {2021})}\BibitemShut {NoStop}%
\bibitem [{\citenamefont {Reed}\ \emph {et~al.}(2010)\citenamefont {Reed}, \citenamefont {Johnson}, \citenamefont {Houck}, \citenamefont {DiCarlo}, \citenamefont {Chow}, \citenamefont {Schuster}, \citenamefont {Frunzio},\ and\ \citenamefont {Schoelkopf}}]{Reed2010}%
  \BibitemOpen
  \bibfield  {author} {\bibinfo {author} {\bibfnamefont {M.~D.}\ \bibnamefont {Reed}}, \bibinfo {author} {\bibfnamefont {B.~R.}\ \bibnamefont {Johnson}}, \bibinfo {author} {\bibfnamefont {A.~A.}\ \bibnamefont {Houck}}, \bibinfo {author} {\bibfnamefont {L.}~\bibnamefont {DiCarlo}}, \bibinfo {author} {\bibfnamefont {J.~M.}\ \bibnamefont {Chow}}, \bibinfo {author} {\bibfnamefont {D.~I.}\ \bibnamefont {Schuster}}, \bibinfo {author} {\bibfnamefont {L.}~\bibnamefont {Frunzio}},\ and\ \bibinfo {author} {\bibfnamefont {R.~J.}\ \bibnamefont {Schoelkopf}},\ }\href {https://doi.org/10.1063/1.3435463} {\bibfield  {journal} {\bibinfo  {journal} {Applied Physics Letters}\ }\textbf {\bibinfo {volume} {96}},\ \bibinfo {pages} {203110} (\bibinfo {year} {2010})}\BibitemShut {NoStop}%
\bibitem [{\citenamefont {Blais}\ \emph {et~al.}(2004)\citenamefont {Blais}, \citenamefont {Huang}, \citenamefont {Wallraff}, \citenamefont {Girvin},\ and\ \citenamefont {Schoelkopf}}]{Blais2004}%
  \BibitemOpen
  \bibfield  {author} {\bibinfo {author} {\bibfnamefont {A.}~\bibnamefont {Blais}}, \bibinfo {author} {\bibfnamefont {R.-S.}\ \bibnamefont {Huang}}, \bibinfo {author} {\bibfnamefont {A.}~\bibnamefont {Wallraff}}, \bibinfo {author} {\bibfnamefont {S.~M.}\ \bibnamefont {Girvin}},\ and\ \bibinfo {author} {\bibfnamefont {R.~J.}\ \bibnamefont {Schoelkopf}},\ }\href {https://doi.org/10.1103/PhysRevA.69.062320} {\bibfield  {journal} {\bibinfo  {journal} {Phys. Rev. A}\ }\textbf {\bibinfo {volume} {69}},\ \bibinfo {pages} {062320} (\bibinfo {year} {2004})}\BibitemShut {NoStop}%
\bibitem [{\citenamefont {Sank}\ \emph {et~al.}(2016)\citenamefont {Sank}, \citenamefont {Chen}, \citenamefont {Khezri}, \citenamefont {Kelly}, \citenamefont {Barends}, \citenamefont {Campbell}, \citenamefont {Chen}, \citenamefont {Chiaro}, \citenamefont {Dunsworth}, \citenamefont {Fowler}, \citenamefont {Jeffrey}, \citenamefont {Lucero}, \citenamefont {Megrant}, \citenamefont {Mutus}, \citenamefont {Neeley}, \citenamefont {Neill}, \citenamefont {O'Malley}, \citenamefont {Quintana}, \citenamefont {Roushan}, \citenamefont {Vainsencher}, \citenamefont {White}, \citenamefont {Wenner}, \citenamefont {Korotkov},\ and\ \citenamefont {Martinis}}]{Sank2016}%
  \BibitemOpen
  \bibfield  {author} {\bibinfo {author} {\bibfnamefont {D.}~\bibnamefont {Sank}}, \bibinfo {author} {\bibfnamefont {Z.}~\bibnamefont {Chen}}, \bibinfo {author} {\bibfnamefont {M.}~\bibnamefont {Khezri}}, \bibinfo {author} {\bibfnamefont {J.}~\bibnamefont {Kelly}}, \bibinfo {author} {\bibfnamefont {R.}~\bibnamefont {Barends}}, \bibinfo {author} {\bibfnamefont {B.}~\bibnamefont {Campbell}}, \bibinfo {author} {\bibfnamefont {Y.}~\bibnamefont {Chen}}, \bibinfo {author} {\bibfnamefont {B.}~\bibnamefont {Chiaro}}, \bibinfo {author} {\bibfnamefont {A.}~\bibnamefont {Dunsworth}}, \bibinfo {author} {\bibfnamefont {A.}~\bibnamefont {Fowler}}, \bibinfo {author} {\bibfnamefont {E.}~\bibnamefont {Jeffrey}}, \bibinfo {author} {\bibfnamefont {E.}~\bibnamefont {Lucero}}, \bibinfo {author} {\bibfnamefont {A.}~\bibnamefont {Megrant}}, \bibinfo {author} {\bibfnamefont {J.}~\bibnamefont {Mutus}}, \bibinfo {author} {\bibfnamefont {M.}~\bibnamefont {Neeley}}, \bibinfo {author} {\bibfnamefont {C.}~\bibnamefont {Neill}}, \bibinfo
  {author} {\bibfnamefont {P.~J.~J.}\ \bibnamefont {O'Malley}}, \bibinfo {author} {\bibfnamefont {C.}~\bibnamefont {Quintana}}, \bibinfo {author} {\bibfnamefont {P.}~\bibnamefont {Roushan}}, \bibinfo {author} {\bibfnamefont {A.}~\bibnamefont {Vainsencher}}, \bibinfo {author} {\bibfnamefont {T.}~\bibnamefont {White}}, \bibinfo {author} {\bibfnamefont {J.}~\bibnamefont {Wenner}}, \bibinfo {author} {\bibfnamefont {A.~N.}\ \bibnamefont {Korotkov}},\ and\ \bibinfo {author} {\bibfnamefont {J.~M.}\ \bibnamefont {Martinis}},\ }\href {https://doi.org/10.1103/PhysRevLett.117.190503} {\bibfield  {journal} {\bibinfo  {journal} {Phys. Rev. Lett.}\ }\textbf {\bibinfo {volume} {117}},\ \bibinfo {pages} {190503} (\bibinfo {year} {2016})}\BibitemShut {NoStop}%
\bibitem [{\citenamefont {Shillito}\ \emph {et~al.}(2022)\citenamefont {Shillito}, \citenamefont {Petrescu}, \citenamefont {Cohen}, \citenamefont {Beall}, \citenamefont {Hauru}, \citenamefont {Ganahl}, \citenamefont {Lewis}, \citenamefont {Vidal},\ and\ \citenamefont {Blais}}]{Shillito2022}%
  \BibitemOpen
  \bibfield  {author} {\bibinfo {author} {\bibfnamefont {R.}~\bibnamefont {Shillito}}, \bibinfo {author} {\bibfnamefont {A.}~\bibnamefont {Petrescu}}, \bibinfo {author} {\bibfnamefont {J.}~\bibnamefont {Cohen}}, \bibinfo {author} {\bibfnamefont {J.}~\bibnamefont {Beall}}, \bibinfo {author} {\bibfnamefont {M.}~\bibnamefont {Hauru}}, \bibinfo {author} {\bibfnamefont {M.}~\bibnamefont {Ganahl}}, \bibinfo {author} {\bibfnamefont {A.~G.}\ \bibnamefont {Lewis}}, \bibinfo {author} {\bibfnamefont {G.}~\bibnamefont {Vidal}},\ and\ \bibinfo {author} {\bibfnamefont {A.}~\bibnamefont {Blais}},\ }\href {https://doi.org/10.1103/PhysRevApplied.18.034031} {\bibfield  {journal} {\bibinfo  {journal} {Phys. Rev. Appl.}\ }\textbf {\bibinfo {volume} {18}},\ \bibinfo {pages} {034031} (\bibinfo {year} {2022})}\BibitemShut {NoStop}%
\bibitem [{\citenamefont {Khezri}\ \emph {et~al.}(2023)\citenamefont {Khezri}, \citenamefont {Opremcak}, \citenamefont {Chen}, \citenamefont {Miao}, \citenamefont {McEwen}, \citenamefont {Bengtsson}, \citenamefont {White}, \citenamefont {Naaman}, \citenamefont {Sank}, \citenamefont {Korotkov}, \citenamefont {Chen},\ and\ \citenamefont {Smelyanskiy}}]{Khezri2023}%
  \BibitemOpen
  \bibfield  {author} {\bibinfo {author} {\bibfnamefont {M.}~\bibnamefont {Khezri}}, \bibinfo {author} {\bibfnamefont {A.}~\bibnamefont {Opremcak}}, \bibinfo {author} {\bibfnamefont {Z.}~\bibnamefont {Chen}}, \bibinfo {author} {\bibfnamefont {K.~C.}\ \bibnamefont {Miao}}, \bibinfo {author} {\bibfnamefont {M.}~\bibnamefont {McEwen}}, \bibinfo {author} {\bibfnamefont {A.}~\bibnamefont {Bengtsson}}, \bibinfo {author} {\bibfnamefont {T.}~\bibnamefont {White}}, \bibinfo {author} {\bibfnamefont {O.}~\bibnamefont {Naaman}}, \bibinfo {author} {\bibfnamefont {D.}~\bibnamefont {Sank}}, \bibinfo {author} {\bibfnamefont {A.~N.}\ \bibnamefont {Korotkov}}, \bibinfo {author} {\bibfnamefont {Y.}~\bibnamefont {Chen}},\ and\ \bibinfo {author} {\bibfnamefont {V.}~\bibnamefont {Smelyanskiy}},\ }\href {https://doi.org/10.1103/PhysRevApplied.20.054008} {\bibfield  {journal} {\bibinfo  {journal} {Phys. Rev. Appl.}\ }\textbf {\bibinfo {volume} {20}},\ \bibinfo {pages} {054008} (\bibinfo {year} {2023})}\BibitemShut {NoStop}%
\bibitem [{\citenamefont {Nesterov}\ and\ \citenamefont {Pechenezhskiy}(2024)}]{Nesterov2024}%
  \BibitemOpen
  \bibfield  {author} {\bibinfo {author} {\bibfnamefont {K.~N.}\ \bibnamefont {Nesterov}}\ and\ \bibinfo {author} {\bibfnamefont {I.~V.}\ \bibnamefont {Pechenezhskiy}},\ }\href {https://doi.org/10.1103/PhysRevApplied.22.064038} {\bibfield  {journal} {\bibinfo  {journal} {Phys. Rev. Appl.}\ }\textbf {\bibinfo {volume} {22}},\ \bibinfo {pages} {064038} (\bibinfo {year} {2024})}\BibitemShut {NoStop}%
\bibitem [{\citenamefont {Dumas}\ \emph {et~al.}(2024)\citenamefont {Dumas}, \citenamefont {Groleau-Par\'e}, \citenamefont {McDonald}, \citenamefont {Mu\~noz Arias}, \citenamefont {Lled\'o}, \citenamefont {D'Anjou},\ and\ \citenamefont {Blais}}]{Dumas2024}%
  \BibitemOpen
  \bibfield  {author} {\bibinfo {author} {\bibfnamefont {M.~F.}\ \bibnamefont {Dumas}}, \bibinfo {author} {\bibfnamefont {B.}~\bibnamefont {Groleau-Par\'e}}, \bibinfo {author} {\bibfnamefont {A.}~\bibnamefont {McDonald}}, \bibinfo {author} {\bibfnamefont {M.~H.}\ \bibnamefont {Mu\~noz Arias}}, \bibinfo {author} {\bibfnamefont {C.}~\bibnamefont {Lled\'o}}, \bibinfo {author} {\bibfnamefont {B.}~\bibnamefont {D'Anjou}},\ and\ \bibinfo {author} {\bibfnamefont {A.}~\bibnamefont {Blais}},\ }\href {https://doi.org/10.1103/PhysRevX.14.041023} {\bibfield  {journal} {\bibinfo  {journal} {Phys. Rev. X}\ }\textbf {\bibinfo {volume} {14}},\ \bibinfo {pages} {041023} (\bibinfo {year} {2024})}\BibitemShut {NoStop}%
\bibitem [{\citenamefont {Vool}\ \emph {et~al.}(2014)\citenamefont {Vool}, \citenamefont {Pop}, \citenamefont {Sliwa}, \citenamefont {Abdo}, \citenamefont {Wang}, \citenamefont {Brecht}, \citenamefont {Gao}, \citenamefont {Shankar}, \citenamefont {Hatridge}, \citenamefont {Catelani}, \citenamefont {Mirrahimi}, \citenamefont {Frunzio}, \citenamefont {Schoelkopf}, \citenamefont {Glazman},\ and\ \citenamefont {Devoret}}]{Vool2014}%
  \BibitemOpen
  \bibfield  {author} {\bibinfo {author} {\bibfnamefont {U.}~\bibnamefont {Vool}}, \bibinfo {author} {\bibfnamefont {I.~M.}\ \bibnamefont {Pop}}, \bibinfo {author} {\bibfnamefont {K.}~\bibnamefont {Sliwa}}, \bibinfo {author} {\bibfnamefont {B.}~\bibnamefont {Abdo}}, \bibinfo {author} {\bibfnamefont {C.}~\bibnamefont {Wang}}, \bibinfo {author} {\bibfnamefont {T.}~\bibnamefont {Brecht}}, \bibinfo {author} {\bibfnamefont {Y.~Y.}\ \bibnamefont {Gao}}, \bibinfo {author} {\bibfnamefont {S.}~\bibnamefont {Shankar}}, \bibinfo {author} {\bibfnamefont {M.}~\bibnamefont {Hatridge}}, \bibinfo {author} {\bibfnamefont {G.}~\bibnamefont {Catelani}}, \bibinfo {author} {\bibfnamefont {M.}~\bibnamefont {Mirrahimi}}, \bibinfo {author} {\bibfnamefont {L.}~\bibnamefont {Frunzio}}, \bibinfo {author} {\bibfnamefont {R.~J.}\ \bibnamefont {Schoelkopf}}, \bibinfo {author} {\bibfnamefont {L.~I.}\ \bibnamefont {Glazman}},\ and\ \bibinfo {author} {\bibfnamefont {M.~H.}\ \bibnamefont {Devoret}},\ }\href
  {https://doi.org/10.1103/PhysRevLett.113.247001} {\bibfield  {journal} {\bibinfo  {journal} {Phys. Rev. Lett.}\ }\textbf {\bibinfo {volume} {113}},\ \bibinfo {pages} {247001} (\bibinfo {year} {2014})}\BibitemShut {NoStop}%
\bibitem [{\citenamefont {Kou}\ \emph {et~al.}(2018)\citenamefont {Kou}, \citenamefont {Smith}, \citenamefont {Vool}, \citenamefont {Pop}, \citenamefont {Sliwa}, \citenamefont {Hatridge}, \citenamefont {Frunzio},\ and\ \citenamefont {Devoret}}]{Kou2018}%
  \BibitemOpen
  \bibfield  {author} {\bibinfo {author} {\bibfnamefont {A.}~\bibnamefont {Kou}}, \bibinfo {author} {\bibfnamefont {W.~C.}\ \bibnamefont {Smith}}, \bibinfo {author} {\bibfnamefont {U.}~\bibnamefont {Vool}}, \bibinfo {author} {\bibfnamefont {I.~M.}\ \bibnamefont {Pop}}, \bibinfo {author} {\bibfnamefont {K.~M.}\ \bibnamefont {Sliwa}}, \bibinfo {author} {\bibfnamefont {M.}~\bibnamefont {Hatridge}}, \bibinfo {author} {\bibfnamefont {L.}~\bibnamefont {Frunzio}},\ and\ \bibinfo {author} {\bibfnamefont {M.~H.}\ \bibnamefont {Devoret}},\ }\href {https://doi.org/10.1103/PhysRevApplied.9.064022} {\bibfield  {journal} {\bibinfo  {journal} {Phys. Rev. Appl.}\ }\textbf {\bibinfo {volume} {9}},\ \bibinfo {pages} {064022} (\bibinfo {year} {2018})}\BibitemShut {NoStop}%
\bibitem [{\citenamefont {Ficheux}\ \emph {et~al.}(2021)\citenamefont {Ficheux}, \citenamefont {Nguyen}, \citenamefont {Somoroff}, \citenamefont {Xiong}, \citenamefont {Nesterov}, \citenamefont {Vavilov},\ and\ \citenamefont {Manucharyan}}]{Ficheux2021}%
  \BibitemOpen
  \bibfield  {author} {\bibinfo {author} {\bibfnamefont {Q.}~\bibnamefont {Ficheux}}, \bibinfo {author} {\bibfnamefont {L.~B.}\ \bibnamefont {Nguyen}}, \bibinfo {author} {\bibfnamefont {A.}~\bibnamefont {Somoroff}}, \bibinfo {author} {\bibfnamefont {H.}~\bibnamefont {Xiong}}, \bibinfo {author} {\bibfnamefont {K.~N.}\ \bibnamefont {Nesterov}}, \bibinfo {author} {\bibfnamefont {M.~G.}\ \bibnamefont {Vavilov}},\ and\ \bibinfo {author} {\bibfnamefont {V.~E.}\ \bibnamefont {Manucharyan}},\ }\href {https://doi.org/10.1103/PhysRevX.11.021026} {\bibfield  {journal} {\bibinfo  {journal} {Phys. Rev. X}\ }\textbf {\bibinfo {volume} {11}},\ \bibinfo {pages} {021026} (\bibinfo {year} {2021})}\BibitemShut {NoStop}%
\bibitem [{\citenamefont {Gusenkova}\ \emph {et~al.}(2021)\citenamefont {Gusenkova}, \citenamefont {Spiecker}, \citenamefont {Gebauer}, \citenamefont {Willsch}, \citenamefont {Willsch}, \citenamefont {Valenti}, \citenamefont {Karcher}, \citenamefont {Gr\"unhaupt}, \citenamefont {Takmakov}, \citenamefont {Winkel}, \citenamefont {Rieger}, \citenamefont {Ustinov}, \citenamefont {Roch}, \citenamefont {Wernsdorfer}, \citenamefont {Michielsen}, \citenamefont {Sander},\ and\ \citenamefont {Pop}}]{Gusenkova2021}%
  \BibitemOpen
  \bibfield  {author} {\bibinfo {author} {\bibfnamefont {D.}~\bibnamefont {Gusenkova}}, \bibinfo {author} {\bibfnamefont {M.}~\bibnamefont {Spiecker}}, \bibinfo {author} {\bibfnamefont {R.}~\bibnamefont {Gebauer}}, \bibinfo {author} {\bibfnamefont {M.}~\bibnamefont {Willsch}}, \bibinfo {author} {\bibfnamefont {D.}~\bibnamefont {Willsch}}, \bibinfo {author} {\bibfnamefont {F.}~\bibnamefont {Valenti}}, \bibinfo {author} {\bibfnamefont {N.}~\bibnamefont {Karcher}}, \bibinfo {author} {\bibfnamefont {L.}~\bibnamefont {Gr\"unhaupt}}, \bibinfo {author} {\bibfnamefont {I.}~\bibnamefont {Takmakov}}, \bibinfo {author} {\bibfnamefont {P.}~\bibnamefont {Winkel}}, \bibinfo {author} {\bibfnamefont {D.}~\bibnamefont {Rieger}}, \bibinfo {author} {\bibfnamefont {A.~V.}\ \bibnamefont {Ustinov}}, \bibinfo {author} {\bibfnamefont {N.}~\bibnamefont {Roch}}, \bibinfo {author} {\bibfnamefont {W.}~\bibnamefont {Wernsdorfer}}, \bibinfo {author} {\bibfnamefont {K.}~\bibnamefont {Michielsen}}, \bibinfo {author} {\bibfnamefont
  {O.}~\bibnamefont {Sander}},\ and\ \bibinfo {author} {\bibfnamefont {I.~M.}\ \bibnamefont {Pop}},\ }\href {https://doi.org/10.1103/PhysRevApplied.15.064030} {\bibfield  {journal} {\bibinfo  {journal} {Phys. Rev. Appl.}\ }\textbf {\bibinfo {volume} {15}},\ \bibinfo {pages} {064030} (\bibinfo {year} {2021})}\BibitemShut {NoStop}%
\bibitem [{\citenamefont {Ding}\ \emph {et~al.}(2023)\citenamefont {Ding}, \citenamefont {Hays}, \citenamefont {Sung}, \citenamefont {Kannan}, \citenamefont {An}, \citenamefont {Di~Paolo}, \citenamefont {Karamlou}, \citenamefont {Hazard}, \citenamefont {Azar}, \citenamefont {Kim}, \citenamefont {Niedzielski}, \citenamefont {Melville}, \citenamefont {Schwartz}, \citenamefont {Yoder}, \citenamefont {Orlando}, \citenamefont {Gustavsson}, \citenamefont {Grover}, \citenamefont {Serniak},\ and\ \citenamefont {Oliver}}]{Ding2023}%
  \BibitemOpen
  \bibfield  {author} {\bibinfo {author} {\bibfnamefont {L.}~\bibnamefont {Ding}}, \bibinfo {author} {\bibfnamefont {M.}~\bibnamefont {Hays}}, \bibinfo {author} {\bibfnamefont {Y.}~\bibnamefont {Sung}}, \bibinfo {author} {\bibfnamefont {B.}~\bibnamefont {Kannan}}, \bibinfo {author} {\bibfnamefont {J.}~\bibnamefont {An}}, \bibinfo {author} {\bibfnamefont {A.}~\bibnamefont {Di~Paolo}}, \bibinfo {author} {\bibfnamefont {A.~H.}\ \bibnamefont {Karamlou}}, \bibinfo {author} {\bibfnamefont {T.~M.}\ \bibnamefont {Hazard}}, \bibinfo {author} {\bibfnamefont {K.}~\bibnamefont {Azar}}, \bibinfo {author} {\bibfnamefont {D.~K.}\ \bibnamefont {Kim}}, \bibinfo {author} {\bibfnamefont {B.~M.}\ \bibnamefont {Niedzielski}}, \bibinfo {author} {\bibfnamefont {A.}~\bibnamefont {Melville}}, \bibinfo {author} {\bibfnamefont {M.~E.}\ \bibnamefont {Schwartz}}, \bibinfo {author} {\bibfnamefont {J.~L.}\ \bibnamefont {Yoder}}, \bibinfo {author} {\bibfnamefont {T.~P.}\ \bibnamefont {Orlando}}, \bibinfo {author} {\bibfnamefont
  {S.}~\bibnamefont {Gustavsson}}, \bibinfo {author} {\bibfnamefont {J.~A.}\ \bibnamefont {Grover}}, \bibinfo {author} {\bibfnamefont {K.}~\bibnamefont {Serniak}},\ and\ \bibinfo {author} {\bibfnamefont {W.~D.}\ \bibnamefont {Oliver}},\ }\href {https://doi.org/https://doi.org/10.1103/PhysRevX.13.031035} {\bibfield  {journal} {\bibinfo  {journal} {Phys. Rev. X}\ }\textbf {\bibinfo {volume} {13}},\ \bibinfo {pages} {031035} (\bibinfo {year} {2023})}\BibitemShut {NoStop}%
\bibitem [{\citenamefont {Cottet}\ \emph {et~al.}(2021)\citenamefont {Cottet}, \citenamefont {Xiong}, \citenamefont {Nguyen}, \citenamefont {Lin},\ and\ \citenamefont {Manucharyan}}]{Cottet2021}%
  \BibitemOpen
  \bibfield  {author} {\bibinfo {author} {\bibfnamefont {N.}~\bibnamefont {Cottet}}, \bibinfo {author} {\bibfnamefont {H.}~\bibnamefont {Xiong}}, \bibinfo {author} {\bibfnamefont {L.~B.}\ \bibnamefont {Nguyen}}, \bibinfo {author} {\bibfnamefont {Y.-H.}\ \bibnamefont {Lin}},\ and\ \bibinfo {author} {\bibfnamefont {V.~E.}\ \bibnamefont {Manucharyan}},\ }\href {https://doi.org/https://doi.org/10.1038/s41467-021-26686-x} {\bibfield  {journal} {\bibinfo  {journal} {Nat. Commun.}\ }\textbf {\bibinfo {volume} {12}},\ \bibinfo {pages} {6383} (\bibinfo {year} {2021})}\BibitemShut {NoStop}%
\bibitem [{\citenamefont {Nagourney}\ \emph {et~al.}(1986)\citenamefont {Nagourney}, \citenamefont {Sandberg},\ and\ \citenamefont {Dehmelt}}]{Nagourney1986}%
  \BibitemOpen
  \bibfield  {author} {\bibinfo {author} {\bibfnamefont {W.}~\bibnamefont {Nagourney}}, \bibinfo {author} {\bibfnamefont {J.}~\bibnamefont {Sandberg}},\ and\ \bibinfo {author} {\bibfnamefont {H.}~\bibnamefont {Dehmelt}},\ }\href {https://doi.org/10.1103/PhysRevLett.56.2797} {\bibfield  {journal} {\bibinfo  {journal} {Phys. Rev. Lett.}\ }\textbf {\bibinfo {volume} {56}},\ \bibinfo {pages} {2797} (\bibinfo {year} {1986})}\BibitemShut {NoStop}%
\bibitem [{\citenamefont {Blatt}\ and\ \citenamefont {Zoller}(1988)}]{Blatt1988}%
  \BibitemOpen
  \bibfield  {author} {\bibinfo {author} {\bibfnamefont {R.}~\bibnamefont {Blatt}}\ and\ \bibinfo {author} {\bibfnamefont {P.}~\bibnamefont {Zoller}},\ }\href {https://doi.org/10.1088/0143-0807/9/4/002} {\bibfield  {journal} {\bibinfo  {journal} {Eur. J. Phys.}\ }\textbf {\bibinfo {volume} {9}},\ \bibinfo {pages} {250} (\bibinfo {year} {1988})}\BibitemShut {NoStop}%
\bibitem [{\citenamefont {Monroe}\ \emph {et~al.}(1995)\citenamefont {Monroe}, \citenamefont {Meekhof}, \citenamefont {King}, \citenamefont {Jefferts}, \citenamefont {Itano}, \citenamefont {Wineland},\ and\ \citenamefont {Gould}}]{Monroe1995}%
  \BibitemOpen
  \bibfield  {author} {\bibinfo {author} {\bibfnamefont {C.}~\bibnamefont {Monroe}}, \bibinfo {author} {\bibfnamefont {D.~M.}\ \bibnamefont {Meekhof}}, \bibinfo {author} {\bibfnamefont {B.~E.}\ \bibnamefont {King}}, \bibinfo {author} {\bibfnamefont {S.~R.}\ \bibnamefont {Jefferts}}, \bibinfo {author} {\bibfnamefont {W.~M.}\ \bibnamefont {Itano}}, \bibinfo {author} {\bibfnamefont {D.~J.}\ \bibnamefont {Wineland}},\ and\ \bibinfo {author} {\bibfnamefont {P.}~\bibnamefont {Gould}},\ }\href {https://doi.org/https://doi.org/10.1103/PhysRevLett.75.4011} {\bibfield  {journal} {\bibinfo  {journal} {Phys. Rev. Lett.}\ }\textbf {\bibinfo {volume} {75}},\ \bibinfo {pages} {4011} (\bibinfo {year} {1995})}\BibitemShut {NoStop}%
\bibitem [{\citenamefont {Schmidt}\ \emph {et~al.}(2005)\citenamefont {Schmidt}, \citenamefont {Rosenband}, \citenamefont {Langer}, \citenamefont {Itano}, \citenamefont {Bergquist},\ and\ \citenamefont {Wineland}}]{Schmidt2005}%
  \BibitemOpen
  \bibfield  {author} {\bibinfo {author} {\bibfnamefont {P.~O.}\ \bibnamefont {Schmidt}}, \bibinfo {author} {\bibfnamefont {T.}~\bibnamefont {Rosenband}}, \bibinfo {author} {\bibfnamefont {C.}~\bibnamefont {Langer}}, \bibinfo {author} {\bibfnamefont {W.~M.}\ \bibnamefont {Itano}}, \bibinfo {author} {\bibfnamefont {J.~C.}\ \bibnamefont {Bergquist}},\ and\ \bibinfo {author} {\bibfnamefont {D.~J.}\ \bibnamefont {Wineland}},\ }\href {https://doi.org/10.1126/science.1114375} {\bibfield  {journal} {\bibinfo  {journal} {Science}\ }\textbf {\bibinfo {volume} {309}},\ \bibinfo {pages} {749} (\bibinfo {year} {2005})}\BibitemShut {NoStop}%
\bibitem [{\citenamefont {Hume}\ \emph {et~al.}(2007)\citenamefont {Hume}, \citenamefont {Rosenband},\ and\ \citenamefont {Wineland}}]{Hume2007}%
  \BibitemOpen
  \bibfield  {author} {\bibinfo {author} {\bibfnamefont {D.~B.}\ \bibnamefont {Hume}}, \bibinfo {author} {\bibfnamefont {T.}~\bibnamefont {Rosenband}},\ and\ \bibinfo {author} {\bibfnamefont {D.~J.}\ \bibnamefont {Wineland}},\ }\href {https://doi.org/10.1103/PhysRevLett.99.120502} {\bibfield  {journal} {\bibinfo  {journal} {Phys. Rev. Lett.}\ }\textbf {\bibinfo {volume} {99}},\ \bibinfo {pages} {120502} (\bibinfo {year} {2007})}\BibitemShut {NoStop}%
\bibitem [{\citenamefont {Magnard}\ \emph {et~al.}(2018)\citenamefont {Magnard}, \citenamefont {Kurpiers}, \citenamefont {Royer}, \citenamefont {Walter}, \citenamefont {Besse}, \citenamefont {Gasparinetti}, \citenamefont {Pechal}, \citenamefont {Heinsoo}, \citenamefont {Storz}, \citenamefont {Blais},\ and\ \citenamefont {Wallraff}}]{Magnard2018}%
  \BibitemOpen
  \bibfield  {author} {\bibinfo {author} {\bibfnamefont {P.}~\bibnamefont {Magnard}}, \bibinfo {author} {\bibfnamefont {P.}~\bibnamefont {Kurpiers}}, \bibinfo {author} {\bibfnamefont {B.}~\bibnamefont {Royer}}, \bibinfo {author} {\bibfnamefont {T.}~\bibnamefont {Walter}}, \bibinfo {author} {\bibfnamefont {J.-C.}\ \bibnamefont {Besse}}, \bibinfo {author} {\bibfnamefont {S.}~\bibnamefont {Gasparinetti}}, \bibinfo {author} {\bibfnamefont {M.}~\bibnamefont {Pechal}}, \bibinfo {author} {\bibfnamefont {J.}~\bibnamefont {Heinsoo}}, \bibinfo {author} {\bibfnamefont {S.}~\bibnamefont {Storz}}, \bibinfo {author} {\bibfnamefont {A.}~\bibnamefont {Blais}},\ and\ \bibinfo {author} {\bibfnamefont {A.}~\bibnamefont {Wallraff}},\ }\href {https://doi.org/10.1103/PhysRevLett.121.060502} {\bibfield  {journal} {\bibinfo  {journal} {Phys. Rev. Lett.}\ }\textbf {\bibinfo {volume} {121}},\ \bibinfo {pages} {060502} (\bibinfo {year} {2018})}\BibitemShut {NoStop}%
\bibitem [{\citenamefont {Sunada}\ \emph {et~al.}(2022)\citenamefont {Sunada}, \citenamefont {Kono}, \citenamefont {Ilves}, \citenamefont {Tamate}, \citenamefont {Sugiyama}, \citenamefont {Tabuchi},\ and\ \citenamefont {Nakamura}}]{Sunada2022}%
  \BibitemOpen
  \bibfield  {author} {\bibinfo {author} {\bibfnamefont {Y.}~\bibnamefont {Sunada}}, \bibinfo {author} {\bibfnamefont {S.}~\bibnamefont {Kono}}, \bibinfo {author} {\bibfnamefont {J.}~\bibnamefont {Ilves}}, \bibinfo {author} {\bibfnamefont {S.}~\bibnamefont {Tamate}}, \bibinfo {author} {\bibfnamefont {T.}~\bibnamefont {Sugiyama}}, \bibinfo {author} {\bibfnamefont {Y.}~\bibnamefont {Tabuchi}},\ and\ \bibinfo {author} {\bibfnamefont {Y.}~\bibnamefont {Nakamura}},\ }\href {https://doi.org/10.1103/PhysRevApplied.17.044016} {\bibfield  {journal} {\bibinfo  {journal} {Phys. Rev. Appl.}\ }\textbf {\bibinfo {volume} {17}},\ \bibinfo {pages} {044016} (\bibinfo {year} {2022})}\BibitemShut {NoStop}%
\bibitem [{\citenamefont {Zhang}\ \emph {et~al.}(2021)\citenamefont {Zhang}, \citenamefont {Chakram}, \citenamefont {Roy}, \citenamefont {Earnest}, \citenamefont {Lu}, \citenamefont {Huang}, \citenamefont {Weiss}, \citenamefont {Koch},\ and\ \citenamefont {Schuster}}]{Zhang2021}%
  \BibitemOpen
  \bibfield  {author} {\bibinfo {author} {\bibfnamefont {H.}~\bibnamefont {Zhang}}, \bibinfo {author} {\bibfnamefont {S.}~\bibnamefont {Chakram}}, \bibinfo {author} {\bibfnamefont {T.}~\bibnamefont {Roy}}, \bibinfo {author} {\bibfnamefont {N.}~\bibnamefont {Earnest}}, \bibinfo {author} {\bibfnamefont {Y.}~\bibnamefont {Lu}}, \bibinfo {author} {\bibfnamefont {Z.}~\bibnamefont {Huang}}, \bibinfo {author} {\bibfnamefont {D.~K.}\ \bibnamefont {Weiss}}, \bibinfo {author} {\bibfnamefont {J.}~\bibnamefont {Koch}},\ and\ \bibinfo {author} {\bibfnamefont {D.~I.}\ \bibnamefont {Schuster}},\ }\href {https://doi.org/10.1103/PhysRevX.11.011010} {\bibfield  {journal} {\bibinfo  {journal} {Phys. Rev. X}\ }\textbf {\bibinfo {volume} {11}},\ \bibinfo {pages} {011010} (\bibinfo {year} {2021})}\BibitemShut {NoStop}%
\bibitem [{\citenamefont {Bao}\ \emph {et~al.}(2022)\citenamefont {Bao}, \citenamefont {Deng}, \citenamefont {Ding}, \citenamefont {Gao}, \citenamefont {Gao}, \citenamefont {Huang}, \citenamefont {Jiang}, \citenamefont {Ku}, \citenamefont {Li}, \citenamefont {Ma}, \citenamefont {Ni}, \citenamefont {Qin}, \citenamefont {Song}, \citenamefont {Sun}, \citenamefont {Tang}, \citenamefont {Wang}, \citenamefont {Wu}, \citenamefont {Xia}, \citenamefont {Yu}, \citenamefont {Zhang}, \citenamefont {Zhang}, \citenamefont {Zhang}, \citenamefont {Zhou}, \citenamefont {Zhu}, \citenamefont {Shi}, \citenamefont {Chen}, \citenamefont {Zhao},\ and\ \citenamefont {Deng}}]{Bao2022}%
  \BibitemOpen
  \bibfield  {author} {\bibinfo {author} {\bibfnamefont {F.}~\bibnamefont {Bao}}, \bibinfo {author} {\bibfnamefont {H.}~\bibnamefont {Deng}}, \bibinfo {author} {\bibfnamefont {D.}~\bibnamefont {Ding}}, \bibinfo {author} {\bibfnamefont {R.}~\bibnamefont {Gao}}, \bibinfo {author} {\bibfnamefont {X.}~\bibnamefont {Gao}}, \bibinfo {author} {\bibfnamefont {C.}~\bibnamefont {Huang}}, \bibinfo {author} {\bibfnamefont {X.}~\bibnamefont {Jiang}}, \bibinfo {author} {\bibfnamefont {H.-S.}\ \bibnamefont {Ku}}, \bibinfo {author} {\bibfnamefont {Z.}~\bibnamefont {Li}}, \bibinfo {author} {\bibfnamefont {X.}~\bibnamefont {Ma}}, \bibinfo {author} {\bibfnamefont {X.}~\bibnamefont {Ni}}, \bibinfo {author} {\bibfnamefont {J.}~\bibnamefont {Qin}}, \bibinfo {author} {\bibfnamefont {Z.}~\bibnamefont {Song}}, \bibinfo {author} {\bibfnamefont {H.}~\bibnamefont {Sun}}, \bibinfo {author} {\bibfnamefont {C.}~\bibnamefont {Tang}}, \bibinfo {author} {\bibfnamefont {T.}~\bibnamefont {Wang}}, \bibinfo {author} {\bibfnamefont
  {F.}~\bibnamefont {Wu}}, \bibinfo {author} {\bibfnamefont {T.}~\bibnamefont {Xia}}, \bibinfo {author} {\bibfnamefont {W.}~\bibnamefont {Yu}}, \bibinfo {author} {\bibfnamefont {F.}~\bibnamefont {Zhang}}, \bibinfo {author} {\bibfnamefont {G.}~\bibnamefont {Zhang}}, \bibinfo {author} {\bibfnamefont {X.}~\bibnamefont {Zhang}}, \bibinfo {author} {\bibfnamefont {J.}~\bibnamefont {Zhou}}, \bibinfo {author} {\bibfnamefont {X.}~\bibnamefont {Zhu}}, \bibinfo {author} {\bibfnamefont {Y.}~\bibnamefont {Shi}}, \bibinfo {author} {\bibfnamefont {J.}~\bibnamefont {Chen}}, \bibinfo {author} {\bibfnamefont {H.-H.}\ \bibnamefont {Zhao}},\ and\ \bibinfo {author} {\bibfnamefont {C.}~\bibnamefont {Deng}},\ }\href {https://doi.org/10.1103/PhysRevLett.129.010502} {\bibfield  {journal} {\bibinfo  {journal} {Phys. Rev. Lett.}\ }\textbf {\bibinfo {volume} {129}},\ \bibinfo {pages} {010502} (\bibinfo {year} {2022})}\BibitemShut {NoStop}%
\bibitem [{\citenamefont {Wang}\ \emph {et~al.}(2024{\natexlab{b}})\citenamefont {Wang}, \citenamefont {Wu}, \citenamefont {Wang}, \citenamefont {Ma}, \citenamefont {Zhang}, \citenamefont {Chen}, \citenamefont {Deng}, \citenamefont {Gao}, \citenamefont {Hu}, \citenamefont {Ma}, \citenamefont {Song}, \citenamefont {Xia}, \citenamefont {Ying}, \citenamefont {Zhan}, \citenamefont {Zhao},\ and\ \citenamefont {Deng}}]{Wang2024}%
  \BibitemOpen
  \bibfield  {author} {\bibinfo {author} {\bibfnamefont {T.}~\bibnamefont {Wang}}, \bibinfo {author} {\bibfnamefont {F.}~\bibnamefont {Wu}}, \bibinfo {author} {\bibfnamefont {F.}~\bibnamefont {Wang}}, \bibinfo {author} {\bibfnamefont {X.}~\bibnamefont {Ma}}, \bibinfo {author} {\bibfnamefont {G.}~\bibnamefont {Zhang}}, \bibinfo {author} {\bibfnamefont {J.}~\bibnamefont {Chen}}, \bibinfo {author} {\bibfnamefont {H.}~\bibnamefont {Deng}}, \bibinfo {author} {\bibfnamefont {R.}~\bibnamefont {Gao}}, \bibinfo {author} {\bibfnamefont {R.}~\bibnamefont {Hu}}, \bibinfo {author} {\bibfnamefont {L.}~\bibnamefont {Ma}}, \bibinfo {author} {\bibfnamefont {Z.}~\bibnamefont {Song}}, \bibinfo {author} {\bibfnamefont {T.}~\bibnamefont {Xia}}, \bibinfo {author} {\bibfnamefont {M.}~\bibnamefont {Ying}}, \bibinfo {author} {\bibfnamefont {H.}~\bibnamefont {Zhan}}, \bibinfo {author} {\bibfnamefont {H.-H.}\ \bibnamefont {Zhao}},\ and\ \bibinfo {author} {\bibfnamefont {C.}~\bibnamefont {Deng}},\ }\href
  {https://doi.org/10.1103/PhysRevLett.132.230601} {\bibfield  {journal} {\bibinfo  {journal} {Phys. Rev. Lett.}\ }\textbf {\bibinfo {volume} {132}},\ \bibinfo {pages} {230601} (\bibinfo {year} {2024}{\natexlab{b}})}\BibitemShut {NoStop}%
\bibitem [{\citenamefont {Rol}\ \emph {et~al.}(2020)\citenamefont {Rol}, \citenamefont {Ciorciaro}, \citenamefont {Malinowski}, \citenamefont {Tarasinski}, \citenamefont {Sagastizabal}, \citenamefont {Bultink}, \citenamefont {Salathe}, \citenamefont {Haandbaek}, \citenamefont {Sedivy},\ and\ \citenamefont {DiCarlo}}]{Rol2020}%
  \BibitemOpen
  \bibfield  {author} {\bibinfo {author} {\bibfnamefont {M.~A.}\ \bibnamefont {Rol}}, \bibinfo {author} {\bibfnamefont {L.}~\bibnamefont {Ciorciaro}}, \bibinfo {author} {\bibfnamefont {F.~K.}\ \bibnamefont {Malinowski}}, \bibinfo {author} {\bibfnamefont {B.~M.}\ \bibnamefont {Tarasinski}}, \bibinfo {author} {\bibfnamefont {R.~E.}\ \bibnamefont {Sagastizabal}}, \bibinfo {author} {\bibfnamefont {C.~C.}\ \bibnamefont {Bultink}}, \bibinfo {author} {\bibfnamefont {Y.}~\bibnamefont {Salathe}}, \bibinfo {author} {\bibfnamefont {N.}~\bibnamefont {Haandbaek}}, \bibinfo {author} {\bibfnamefont {J.}~\bibnamefont {Sedivy}},\ and\ \bibinfo {author} {\bibfnamefont {L.}~\bibnamefont {DiCarlo}},\ }\href {https://doi.org/10.1063/1.5133894} {\bibfield  {journal} {\bibinfo  {journal} {Applied Physics Letters}\ }\textbf {\bibinfo {volume} {116}},\ \bibinfo {pages} {054001} (\bibinfo {year} {2020})}\BibitemShut {NoStop}%
\bibitem [{\citenamefont {Sung}\ \emph {et~al.}(2021)\citenamefont {Sung}, \citenamefont {Ding}, \citenamefont {Braum\"uller}, \citenamefont {Veps\"al\"ainen}, \citenamefont {Kannan}, \citenamefont {Kjaergaard}, \citenamefont {Greene}, \citenamefont {Samach}, \citenamefont {McNally}, \citenamefont {Kim}, \citenamefont {Melville}, \citenamefont {Niedzielski}, \citenamefont {Schwartz}, \citenamefont {Yoder}, \citenamefont {Orlando}, \citenamefont {Gustavsson},\ and\ \citenamefont {Oliver}}]{Sung2021}%
  \BibitemOpen
  \bibfield  {author} {\bibinfo {author} {\bibfnamefont {Y.}~\bibnamefont {Sung}}, \bibinfo {author} {\bibfnamefont {L.}~\bibnamefont {Ding}}, \bibinfo {author} {\bibfnamefont {J.}~\bibnamefont {Braum\"uller}}, \bibinfo {author} {\bibfnamefont {A.}~\bibnamefont {Veps\"al\"ainen}}, \bibinfo {author} {\bibfnamefont {B.}~\bibnamefont {Kannan}}, \bibinfo {author} {\bibfnamefont {M.}~\bibnamefont {Kjaergaard}}, \bibinfo {author} {\bibfnamefont {A.}~\bibnamefont {Greene}}, \bibinfo {author} {\bibfnamefont {G.~O.}\ \bibnamefont {Samach}}, \bibinfo {author} {\bibfnamefont {C.}~\bibnamefont {McNally}}, \bibinfo {author} {\bibfnamefont {D.}~\bibnamefont {Kim}}, \bibinfo {author} {\bibfnamefont {A.}~\bibnamefont {Melville}}, \bibinfo {author} {\bibfnamefont {B.~M.}\ \bibnamefont {Niedzielski}}, \bibinfo {author} {\bibfnamefont {M.~E.}\ \bibnamefont {Schwartz}}, \bibinfo {author} {\bibfnamefont {J.~L.}\ \bibnamefont {Yoder}}, \bibinfo {author} {\bibfnamefont {T.~P.}\ \bibnamefont {Orlando}}, \bibinfo {author}
  {\bibfnamefont {S.}~\bibnamefont {Gustavsson}},\ and\ \bibinfo {author} {\bibfnamefont {W.~D.}\ \bibnamefont {Oliver}},\ }\href {https://doi.org/10.1103/PhysRevX.11.021058} {\bibfield  {journal} {\bibinfo  {journal} {Phys. Rev. X}\ }\textbf {\bibinfo {volume} {11}},\ \bibinfo {pages} {021058} (\bibinfo {year} {2021})}\BibitemShut {NoStop}%
\bibitem [{\citenamefont {Li}\ \emph {et~al.}(2024)\citenamefont {Li}, \citenamefont {Kubo}, \citenamefont {Ho}, \citenamefont {Yan}, \citenamefont {Nakamura},\ and\ \citenamefont {Goto}}]{Li2024}%
  \BibitemOpen
  \bibfield  {author} {\bibinfo {author} {\bibfnamefont {R.}~\bibnamefont {Li}}, \bibinfo {author} {\bibfnamefont {K.}~\bibnamefont {Kubo}}, \bibinfo {author} {\bibfnamefont {Y.}~\bibnamefont {Ho}}, \bibinfo {author} {\bibfnamefont {Z.}~\bibnamefont {Yan}}, \bibinfo {author} {\bibfnamefont {Y.}~\bibnamefont {Nakamura}},\ and\ \bibinfo {author} {\bibfnamefont {H.}~\bibnamefont {Goto}},\ }\href {https://doi.org/10.1103/PhysRevX.14.041050} {\bibfield  {journal} {\bibinfo  {journal} {Phys. Rev. X}\ }\textbf {\bibinfo {volume} {14}},\ \bibinfo {pages} {041050} (\bibinfo {year} {2024})}\BibitemShut {NoStop}%
\bibitem [{\citenamefont {Yan}\ \emph {et~al.}(2023)\citenamefont {Yan}, \citenamefont {Wu}, \citenamefont {Lingenfelter}, \citenamefont {Joshi}, \citenamefont {Andersson}, \citenamefont {Conner}, \citenamefont {Chou}, \citenamefont {Grebel}, \citenamefont {Miller}, \citenamefont {Povey}, \citenamefont {Qiao}, \citenamefont {Clerk},\ and\ \citenamefont {Cleland}}]{Yan2023}%
  \BibitemOpen
  \bibfield  {author} {\bibinfo {author} {\bibfnamefont {H.}~\bibnamefont {Yan}}, \bibinfo {author} {\bibfnamefont {X.}~\bibnamefont {Wu}}, \bibinfo {author} {\bibfnamefont {A.}~\bibnamefont {Lingenfelter}}, \bibinfo {author} {\bibfnamefont {Y.~J.}\ \bibnamefont {Joshi}}, \bibinfo {author} {\bibfnamefont {G.}~\bibnamefont {Andersson}}, \bibinfo {author} {\bibfnamefont {C.~R.}\ \bibnamefont {Conner}}, \bibinfo {author} {\bibfnamefont {M.-H.}\ \bibnamefont {Chou}}, \bibinfo {author} {\bibfnamefont {J.}~\bibnamefont {Grebel}}, \bibinfo {author} {\bibfnamefont {J.~M.}\ \bibnamefont {Miller}}, \bibinfo {author} {\bibfnamefont {R.~G.}\ \bibnamefont {Povey}}, \bibinfo {author} {\bibfnamefont {H.}~\bibnamefont {Qiao}}, \bibinfo {author} {\bibfnamefont {A.~A.}\ \bibnamefont {Clerk}},\ and\ \bibinfo {author} {\bibfnamefont {A.~N.}\ \bibnamefont {Cleland}},\ }\href {https://doi.org/10.1063/5.0161893} {\bibfield  {journal} {\bibinfo  {journal} {Applied Physics Letters}\ }\textbf {\bibinfo {volume} {123}},\ \bibinfo
  {pages} {134001} (\bibinfo {year} {2023})}\BibitemShut {NoStop}%
\bibitem [{\citenamefont {Park}\ \emph {et~al.}(2024)\citenamefont {Park}, \citenamefont {Choi}, \citenamefont {Kim}, \citenamefont {Jo}, \citenamefont {Lee}, \citenamefont {Kim}, \citenamefont {Park}, \citenamefont {Lee},\ and\ \citenamefont {Hahn}}]{Park2024}%
  \BibitemOpen
  \bibfield  {author} {\bibinfo {author} {\bibfnamefont {S.~H.}\ \bibnamefont {Park}}, \bibinfo {author} {\bibfnamefont {G.}~\bibnamefont {Choi}}, \bibinfo {author} {\bibfnamefont {G.}~\bibnamefont {Kim}}, \bibinfo {author} {\bibfnamefont {J.}~\bibnamefont {Jo}}, \bibinfo {author} {\bibfnamefont {B.}~\bibnamefont {Lee}}, \bibinfo {author} {\bibfnamefont {G.}~\bibnamefont {Kim}}, \bibinfo {author} {\bibfnamefont {K.}~\bibnamefont {Park}}, \bibinfo {author} {\bibfnamefont {Y.-H.}\ \bibnamefont {Lee}},\ and\ \bibinfo {author} {\bibfnamefont {S.}~\bibnamefont {Hahn}},\ }\href {https://doi.org/https://doi.org/10.1063/5.0182642} {\bibfield  {journal} {\bibinfo  {journal} {Appl. Phys. Lett.}\ }\textbf {\bibinfo {volume} {124}},\ \bibinfo {pages} {044003} (\bibinfo {year} {2024})}\BibitemShut {NoStop}%
\bibitem [{\citenamefont {Paik}\ and\ \citenamefont {Osborn}(2010)}]{Paik2010}%
  \BibitemOpen
  \bibfield  {author} {\bibinfo {author} {\bibfnamefont {H.}~\bibnamefont {Paik}}\ and\ \bibinfo {author} {\bibfnamefont {K.~D.}\ \bibnamefont {Osborn}},\ }\href {https://doi.org/10.1063/1.3309703} {\bibfield  {journal} {\bibinfo  {journal} {Applied Physics Letters}\ }\textbf {\bibinfo {volume} {96}},\ \bibinfo {pages} {072505} (\bibinfo {year} {2010})}\BibitemShut {NoStop}%
\bibitem [{\citenamefont {Cho}\ \emph {et~al.}(2013)\citenamefont {Cho}, \citenamefont {Patel}, \citenamefont {Podkaminer}, \citenamefont {Gao}, \citenamefont {Folkman}, \citenamefont {Bark}, \citenamefont {Lee}, \citenamefont {Zhang}, \citenamefont {Pan}, \citenamefont {McDermott},\ and\ \citenamefont {Eom}}]{Cho2013}%
  \BibitemOpen
  \bibfield  {author} {\bibinfo {author} {\bibfnamefont {K.-H.}\ \bibnamefont {Cho}}, \bibinfo {author} {\bibfnamefont {U.}~\bibnamefont {Patel}}, \bibinfo {author} {\bibfnamefont {J.}~\bibnamefont {Podkaminer}}, \bibinfo {author} {\bibfnamefont {Y.}~\bibnamefont {Gao}}, \bibinfo {author} {\bibfnamefont {C.~M.}\ \bibnamefont {Folkman}}, \bibinfo {author} {\bibfnamefont {C.~W.}\ \bibnamefont {Bark}}, \bibinfo {author} {\bibfnamefont {S.}~\bibnamefont {Lee}}, \bibinfo {author} {\bibfnamefont {Y.}~\bibnamefont {Zhang}}, \bibinfo {author} {\bibfnamefont {X.~Q.}\ \bibnamefont {Pan}}, \bibinfo {author} {\bibfnamefont {R.}~\bibnamefont {McDermott}},\ and\ \bibinfo {author} {\bibfnamefont {C.~B.}\ \bibnamefont {Eom}},\ }\href {https://doi.org/10.1063/1.4822436} {\bibfield  {journal} {\bibinfo  {journal} {APL Materials}\ }\textbf {\bibinfo {volume} {1}},\ \bibinfo {pages} {042115} (\bibinfo {year} {2013})}\BibitemShut {NoStop}%
\bibitem [{\citenamefont {Deng}\ \emph {et~al.}(2014)\citenamefont {Deng}, \citenamefont {Otto},\ and\ \citenamefont {Lupascu}}]{Deng2014}%
  \BibitemOpen
  \bibfield  {author} {\bibinfo {author} {\bibfnamefont {C.}~\bibnamefont {Deng}}, \bibinfo {author} {\bibfnamefont {M.}~\bibnamefont {Otto}},\ and\ \bibinfo {author} {\bibfnamefont {A.}~\bibnamefont {Lupascu}},\ }\href {https://doi.org/10.1063/1.4863686} {\bibfield  {journal} {\bibinfo  {journal} {Applied Physics Letters}\ }\textbf {\bibinfo {volume} {104}},\ \bibinfo {pages} {043506} (\bibinfo {year} {2014})}\BibitemShut {NoStop}%
\bibitem [{\citenamefont {Zotova}\ \emph {et~al.}(2023)\citenamefont {Zotova}, \citenamefont {Wang}, \citenamefont {Semenov}, \citenamefont {Zhou}, \citenamefont {Khrapach}, \citenamefont {Tomonaga}, \citenamefont {Astafiev},\ and\ \citenamefont {Tsai}}]{Zotova2023}%
  \BibitemOpen
  \bibfield  {author} {\bibinfo {author} {\bibfnamefont {J.}~\bibnamefont {Zotova}}, \bibinfo {author} {\bibfnamefont {R.}~\bibnamefont {Wang}}, \bibinfo {author} {\bibfnamefont {A.}~\bibnamefont {Semenov}}, \bibinfo {author} {\bibfnamefont {Y.}~\bibnamefont {Zhou}}, \bibinfo {author} {\bibfnamefont {I.}~\bibnamefont {Khrapach}}, \bibinfo {author} {\bibfnamefont {A.}~\bibnamefont {Tomonaga}}, \bibinfo {author} {\bibfnamefont {O.}~\bibnamefont {Astafiev}},\ and\ \bibinfo {author} {\bibfnamefont {J.-S.}\ \bibnamefont {Tsai}},\ }\href {https://doi.org/10.1103/PhysRevApplied.19.044067} {\bibfield  {journal} {\bibinfo  {journal} {Phys. Rev. Appl.}\ }\textbf {\bibinfo {volume} {19}},\ \bibinfo {pages} {044067} (\bibinfo {year} {2023})}\BibitemShut {NoStop}%
\bibitem [{\citenamefont {Zotova}\ \emph {et~al.}(2024)\citenamefont {Zotova}, \citenamefont {Sanduleanu}, \citenamefont {Fedorov}, \citenamefont {Wang}, \citenamefont {Tsai},\ and\ \citenamefont {Astafiev}}]{Zotova2024}%
  \BibitemOpen
  \bibfield  {author} {\bibinfo {author} {\bibfnamefont {J.}~\bibnamefont {Zotova}}, \bibinfo {author} {\bibfnamefont {S.}~\bibnamefont {Sanduleanu}}, \bibinfo {author} {\bibfnamefont {G.}~\bibnamefont {Fedorov}}, \bibinfo {author} {\bibfnamefont {R.}~\bibnamefont {Wang}}, \bibinfo {author} {\bibfnamefont {J.~S.}\ \bibnamefont {Tsai}},\ and\ \bibinfo {author} {\bibfnamefont {O.}~\bibnamefont {Astafiev}},\ }\href {https://doi.org/https://doi.org/10.1063/5.0194276} {\bibfield  {journal} {\bibinfo  {journal} {Appl. Phys. Lett.}\ }\textbf {\bibinfo {volume} {124}},\ \bibinfo {pages} {102601} (\bibinfo {year} {2024})}\BibitemShut {NoStop}%
\bibitem [{\citenamefont {Geerlings}\ \emph {et~al.}(2012)\citenamefont {Geerlings}, \citenamefont {Shankar}, \citenamefont {Edwards}, \citenamefont {Frunzio}, \citenamefont {Schoelkopf},\ and\ \citenamefont {Devoret}}]{Geerlings2012}%
  \BibitemOpen
  \bibfield  {author} {\bibinfo {author} {\bibfnamefont {K.}~\bibnamefont {Geerlings}}, \bibinfo {author} {\bibfnamefont {S.}~\bibnamefont {Shankar}}, \bibinfo {author} {\bibfnamefont {E.}~\bibnamefont {Edwards}}, \bibinfo {author} {\bibfnamefont {L.}~\bibnamefont {Frunzio}}, \bibinfo {author} {\bibfnamefont {R.~J.}\ \bibnamefont {Schoelkopf}},\ and\ \bibinfo {author} {\bibfnamefont {M.~H.}\ \bibnamefont {Devoret}},\ }\href {https://doi.org/10.1063/1.4710520} {\bibfield  {journal} {\bibinfo  {journal} {Applied Physics Letters}\ }\textbf {\bibinfo {volume} {100}},\ \bibinfo {pages} {192601} (\bibinfo {year} {2012})}\BibitemShut {NoStop}%
\bibitem [{\citenamefont {Jiang}\ \emph {et~al.}(2022)\citenamefont {Jiang}, \citenamefont {Li}, \citenamefont {Guo}, \citenamefont {Xu}, \citenamefont {Wei}, \citenamefont {Zhang}, \citenamefont {Zhou}, \citenamefont {Sheng}, \citenamefont {Cao}, \citenamefont {Sun},\ and\ \citenamefont {Wu}}]{Jiang2022}%
  \BibitemOpen
  \bibfield  {author} {\bibinfo {author} {\bibfnamefont {J.}~\bibnamefont {Jiang}}, \bibinfo {author} {\bibfnamefont {Z.}~\bibnamefont {Li}}, \bibinfo {author} {\bibfnamefont {T.}~\bibnamefont {Guo}}, \bibinfo {author} {\bibfnamefont {W.}~\bibnamefont {Xu}}, \bibinfo {author} {\bibfnamefont {X.}~\bibnamefont {Wei}}, \bibinfo {author} {\bibfnamefont {K.}~\bibnamefont {Zhang}}, \bibinfo {author} {\bibfnamefont {T.}~\bibnamefont {Zhou}}, \bibinfo {author} {\bibfnamefont {Y.}~\bibnamefont {Sheng}}, \bibinfo {author} {\bibfnamefont {C.}~\bibnamefont {Cao}}, \bibinfo {author} {\bibfnamefont {G.}~\bibnamefont {Sun}},\ and\ \bibinfo {author} {\bibfnamefont {P.}~\bibnamefont {Wu}},\ }\href {https://doi.org/10.1063/5.0128964} {\bibfield  {journal} {\bibinfo  {journal} {Applied Physics Letters}\ }\textbf {\bibinfo {volume} {121}},\ \bibinfo {pages} {254001} (\bibinfo {year} {2022})}\BibitemShut {NoStop}%
\bibitem [{com(2022)}]{comsol}%
  \BibitemOpen
  \href@noop {} {\bibinfo {title} {{COMSOL Multiphysics} 6.1}} (\bibinfo {year} {2022}),\ \bibinfo {note} {\url{www.comsol.com}. COMSOL AB, Stockholm, Sweden.}\BibitemShut {Stop}%
\bibitem [{\citenamefont {Margraf}(2022)}]{qucsstudio}%
  \BibitemOpen
  \bibfield  {author} {\bibinfo {author} {\bibfnamefont {M.}~\bibnamefont {Margraf}},\ }\href@noop {} {\bibinfo {title} {{QucsStudio} 4.3.1}} (\bibinfo {year} {2022}),\ \bibinfo {note} {\url{https://qucsstudio.de}}\BibitemShut {NoStop}%
\bibitem [{\citenamefont {Martinis}\ \emph {et~al.}(2009)\citenamefont {Martinis}, \citenamefont {Ansmann},\ and\ \citenamefont {Aumentado}}]{Martinis2009}%
  \BibitemOpen
  \bibfield  {author} {\bibinfo {author} {\bibfnamefont {J.~M.}\ \bibnamefont {Martinis}}, \bibinfo {author} {\bibfnamefont {M.}~\bibnamefont {Ansmann}},\ and\ \bibinfo {author} {\bibfnamefont {J.}~\bibnamefont {Aumentado}},\ }\href {https://doi.org/10.1103/PhysRevLett.103.097002} {\bibfield  {journal} {\bibinfo  {journal} {Phys. Rev. Lett.}\ }\textbf {\bibinfo {volume} {103}},\ \bibinfo {pages} {097002} (\bibinfo {year} {2009})}\BibitemShut {NoStop}%
\bibitem [{\citenamefont {Pop}\ \emph {et~al.}(2014)\citenamefont {Pop}, \citenamefont {Geerlings}, \citenamefont {Catelani}, \citenamefont {Schoelkopf}, \citenamefont {Glazman},\ and\ \citenamefont {Devoret}}]{Pop2014}%
  \BibitemOpen
  \bibfield  {author} {\bibinfo {author} {\bibfnamefont {I.~M.}\ \bibnamefont {Pop}}, \bibinfo {author} {\bibfnamefont {K.}~\bibnamefont {Geerlings}}, \bibinfo {author} {\bibfnamefont {G.}~\bibnamefont {Catelani}}, \bibinfo {author} {\bibfnamefont {R.~J.}\ \bibnamefont {Schoelkopf}}, \bibinfo {author} {\bibfnamefont {L.~I.}\ \bibnamefont {Glazman}},\ and\ \bibinfo {author} {\bibfnamefont {M.~H.}\ \bibnamefont {Devoret}},\ }\href@noop {} {\bibfield  {journal} {\bibinfo  {journal} {Nature}\ }\textbf {\bibinfo {volume} {508}},\ \bibinfo {pages} {369} (\bibinfo {year} {2014})}\BibitemShut {NoStop}%
\bibitem [{\citenamefont {Serniak}\ \emph {et~al.}(2018)\citenamefont {Serniak}, \citenamefont {Hays}, \citenamefont {de~Lange}, \citenamefont {Diamond}, \citenamefont {Shankar}, \citenamefont {Burkhart}, \citenamefont {Frunzio}, \citenamefont {Houzet},\ and\ \citenamefont {Devoret}}]{Serniak2018}%
  \BibitemOpen
  \bibfield  {author} {\bibinfo {author} {\bibfnamefont {K.}~\bibnamefont {Serniak}}, \bibinfo {author} {\bibfnamefont {M.}~\bibnamefont {Hays}}, \bibinfo {author} {\bibfnamefont {G.}~\bibnamefont {de~Lange}}, \bibinfo {author} {\bibfnamefont {S.}~\bibnamefont {Diamond}}, \bibinfo {author} {\bibfnamefont {S.}~\bibnamefont {Shankar}}, \bibinfo {author} {\bibfnamefont {L.~D.}\ \bibnamefont {Burkhart}}, \bibinfo {author} {\bibfnamefont {L.}~\bibnamefont {Frunzio}}, \bibinfo {author} {\bibfnamefont {M.}~\bibnamefont {Houzet}},\ and\ \bibinfo {author} {\bibfnamefont {M.~H.}\ \bibnamefont {Devoret}},\ }\href {https://doi.org/10.1103/PhysRevLett.121.157701} {\bibfield  {journal} {\bibinfo  {journal} {Phys. Rev. Lett.}\ }\textbf {\bibinfo {volume} {121}},\ \bibinfo {pages} {157701} (\bibinfo {year} {2018})}\BibitemShut {NoStop}%
\bibitem [{\citenamefont {Nguyen}\ \emph {et~al.}(2019)\citenamefont {Nguyen}, \citenamefont {Lin}, \citenamefont {Somoroff}, \citenamefont {Mencia}, \citenamefont {Grabon},\ and\ \citenamefont {Manucharyan}}]{Nguyen2019}%
  \BibitemOpen
  \bibfield  {author} {\bibinfo {author} {\bibfnamefont {L.~B.}\ \bibnamefont {Nguyen}}, \bibinfo {author} {\bibfnamefont {Y.-H.}\ \bibnamefont {Lin}}, \bibinfo {author} {\bibfnamefont {A.}~\bibnamefont {Somoroff}}, \bibinfo {author} {\bibfnamefont {R.}~\bibnamefont {Mencia}}, \bibinfo {author} {\bibfnamefont {N.}~\bibnamefont {Grabon}},\ and\ \bibinfo {author} {\bibfnamefont {V.~E.}\ \bibnamefont {Manucharyan}},\ }\href {https://doi.org/10.1103/PhysRevX.9.041041} {\bibfield  {journal} {\bibinfo  {journal} {Phys. Rev. X}\ }\textbf {\bibinfo {volume} {9}},\ \bibinfo {pages} {041041} (\bibinfo {year} {2019})}\BibitemShut {NoStop}%
\bibitem [{\citenamefont {Connolly}\ \emph {et~al.}(2024)\citenamefont {Connolly}, \citenamefont {Kurilovich}, \citenamefont {Diamond}, \citenamefont {Nho}, \citenamefont {B\o{}ttcher}, \citenamefont {Glazman}, \citenamefont {Fatemi},\ and\ \citenamefont {Devoret}}]{Connolly2024}%
  \BibitemOpen
  \bibfield  {author} {\bibinfo {author} {\bibfnamefont {T.}~\bibnamefont {Connolly}}, \bibinfo {author} {\bibfnamefont {P.~D.}\ \bibnamefont {Kurilovich}}, \bibinfo {author} {\bibfnamefont {S.}~\bibnamefont {Diamond}}, \bibinfo {author} {\bibfnamefont {H.}~\bibnamefont {Nho}}, \bibinfo {author} {\bibfnamefont {C.~G.~L.}\ \bibnamefont {B\o{}ttcher}}, \bibinfo {author} {\bibfnamefont {L.~I.}\ \bibnamefont {Glazman}}, \bibinfo {author} {\bibfnamefont {V.}~\bibnamefont {Fatemi}},\ and\ \bibinfo {author} {\bibfnamefont {M.~H.}\ \bibnamefont {Devoret}},\ }\href {https://doi.org/10.1103/PhysRevLett.132.217001} {\bibfield  {journal} {\bibinfo  {journal} {Phys. Rev. Lett.}\ }\textbf {\bibinfo {volume} {132}},\ \bibinfo {pages} {217001} (\bibinfo {year} {2024})}\BibitemShut {NoStop}%
\bibitem [{\citenamefont {Aumentado}\ \emph {et~al.}(2004)\citenamefont {Aumentado}, \citenamefont {Keller}, \citenamefont {Martinis},\ and\ \citenamefont {Devoret}}]{Aumentado2004}%
  \BibitemOpen
  \bibfield  {author} {\bibinfo {author} {\bibfnamefont {J.}~\bibnamefont {Aumentado}}, \bibinfo {author} {\bibfnamefont {M.~W.}\ \bibnamefont {Keller}}, \bibinfo {author} {\bibfnamefont {J.~M.}\ \bibnamefont {Martinis}},\ and\ \bibinfo {author} {\bibfnamefont {M.~H.}\ \bibnamefont {Devoret}},\ }\href {https://doi.org/10.1103/PhysRevLett.92.066802} {\bibfield  {journal} {\bibinfo  {journal} {Phys. Rev. Lett.}\ }\textbf {\bibinfo {volume} {92}},\ \bibinfo {pages} {066802} (\bibinfo {year} {2004})}\BibitemShut {NoStop}%
\bibitem [{\citenamefont {Kalashnikov}\ \emph {et~al.}(2020)\citenamefont {Kalashnikov}, \citenamefont {Hsieh}, \citenamefont {Zhang}, \citenamefont {Lu}, \citenamefont {Kamenov}, \citenamefont {Di~Paolo}, \citenamefont {Blais}, \citenamefont {Gershenson},\ and\ \citenamefont {Bell}}]{Kalashnikov2020}%
  \BibitemOpen
  \bibfield  {author} {\bibinfo {author} {\bibfnamefont {K.}~\bibnamefont {Kalashnikov}}, \bibinfo {author} {\bibfnamefont {W.~T.}\ \bibnamefont {Hsieh}}, \bibinfo {author} {\bibfnamefont {W.}~\bibnamefont {Zhang}}, \bibinfo {author} {\bibfnamefont {W.-S.}\ \bibnamefont {Lu}}, \bibinfo {author} {\bibfnamefont {P.}~\bibnamefont {Kamenov}}, \bibinfo {author} {\bibfnamefont {A.}~\bibnamefont {Di~Paolo}}, \bibinfo {author} {\bibfnamefont {A.}~\bibnamefont {Blais}}, \bibinfo {author} {\bibfnamefont {M.~E.}\ \bibnamefont {Gershenson}},\ and\ \bibinfo {author} {\bibfnamefont {M.}~\bibnamefont {Bell}},\ }\href {https://doi.org/10.1103/PRXQuantum.1.010307} {\bibfield  {journal} {\bibinfo  {journal} {PRX Quantum}\ }\textbf {\bibinfo {volume} {1}},\ \bibinfo {pages} {010307} (\bibinfo {year} {2020})}\BibitemShut {NoStop}%
\bibitem [{\citenamefont {McEwen}\ \emph {et~al.}(2024)\citenamefont {McEwen}, \citenamefont {Miao}, \citenamefont {Atalaya}, \citenamefont {Bilmes}, \citenamefont {Crook}, \citenamefont {Bovaird}, \citenamefont {Kreikebaum}, \citenamefont {Zobrist}, \citenamefont {Jeffrey}, \citenamefont {Ying}, \citenamefont {Bengtsson}, \citenamefont {Chang}, \citenamefont {Dunsworth}, \citenamefont {Kelly}, \citenamefont {Zhang}, \citenamefont {Forati}, \citenamefont {Acharya}, \citenamefont {Iveland}, \citenamefont {Liu}, \citenamefont {Kim}, \citenamefont {Burkett}, \citenamefont {Megrant}, \citenamefont {Chen}, \citenamefont {Neill}, \citenamefont {Sank}, \citenamefont {Devoret},\ and\ \citenamefont {Opremcak}}]{McEwen2024}%
  \BibitemOpen
  \bibfield  {author} {\bibinfo {author} {\bibfnamefont {M.}~\bibnamefont {McEwen}}, \bibinfo {author} {\bibfnamefont {K.~C.}\ \bibnamefont {Miao}}, \bibinfo {author} {\bibfnamefont {J.}~\bibnamefont {Atalaya}}, \bibinfo {author} {\bibfnamefont {A.}~\bibnamefont {Bilmes}}, \bibinfo {author} {\bibfnamefont {A.}~\bibnamefont {Crook}}, \bibinfo {author} {\bibfnamefont {J.}~\bibnamefont {Bovaird}}, \bibinfo {author} {\bibfnamefont {J.~M.}\ \bibnamefont {Kreikebaum}}, \bibinfo {author} {\bibfnamefont {N.}~\bibnamefont {Zobrist}}, \bibinfo {author} {\bibfnamefont {E.}~\bibnamefont {Jeffrey}}, \bibinfo {author} {\bibfnamefont {B.}~\bibnamefont {Ying}}, \bibinfo {author} {\bibfnamefont {A.}~\bibnamefont {Bengtsson}}, \bibinfo {author} {\bibfnamefont {H.-S.}\ \bibnamefont {Chang}}, \bibinfo {author} {\bibfnamefont {A.}~\bibnamefont {Dunsworth}}, \bibinfo {author} {\bibfnamefont {J.}~\bibnamefont {Kelly}}, \bibinfo {author} {\bibfnamefont {Y.}~\bibnamefont {Zhang}}, \bibinfo {author} {\bibfnamefont {E.}~\bibnamefont
  {Forati}}, \bibinfo {author} {\bibfnamefont {R.}~\bibnamefont {Acharya}}, \bibinfo {author} {\bibfnamefont {J.}~\bibnamefont {Iveland}}, \bibinfo {author} {\bibfnamefont {W.}~\bibnamefont {Liu}}, \bibinfo {author} {\bibfnamefont {S.}~\bibnamefont {Kim}}, \bibinfo {author} {\bibfnamefont {B.}~\bibnamefont {Burkett}}, \bibinfo {author} {\bibfnamefont {A.}~\bibnamefont {Megrant}}, \bibinfo {author} {\bibfnamefont {Y.}~\bibnamefont {Chen}}, \bibinfo {author} {\bibfnamefont {C.}~\bibnamefont {Neill}}, \bibinfo {author} {\bibfnamefont {D.}~\bibnamefont {Sank}}, \bibinfo {author} {\bibfnamefont {M.}~\bibnamefont {Devoret}},\ and\ \bibinfo {author} {\bibfnamefont {A.}~\bibnamefont {Opremcak}},\ }\href {https://doi.org/10.1103/PhysRevLett.133.240601} {\bibfield  {journal} {\bibinfo  {journal} {Phys. Rev. Lett.}\ }\textbf {\bibinfo {volume} {133}},\ \bibinfo {pages} {240601} (\bibinfo {year} {2024})}\BibitemShut {NoStop}%
\bibitem [{\citenamefont {Moskalenko}\ \emph {et~al.}(2022)\citenamefont {Moskalenko}, \citenamefont {Simakov}, \citenamefont {Abramov}, \citenamefont {Grigorev}, \citenamefont {Moskalev}, \citenamefont {Pishchimova}, \citenamefont {Smirnov}, \citenamefont {Zikiy}, \citenamefont {Rodionov},\ and\ \citenamefont {Besedin}}]{Moskalenko2022}%
  \BibitemOpen
  \bibfield  {author} {\bibinfo {author} {\bibfnamefont {I.~N.}\ \bibnamefont {Moskalenko}}, \bibinfo {author} {\bibfnamefont {I.~A.}\ \bibnamefont {Simakov}}, \bibinfo {author} {\bibfnamefont {N.~N.}\ \bibnamefont {Abramov}}, \bibinfo {author} {\bibfnamefont {A.~A.}\ \bibnamefont {Grigorev}}, \bibinfo {author} {\bibfnamefont {D.~O.}\ \bibnamefont {Moskalev}}, \bibinfo {author} {\bibfnamefont {A.~A.}\ \bibnamefont {Pishchimova}}, \bibinfo {author} {\bibfnamefont {N.~S.}\ \bibnamefont {Smirnov}}, \bibinfo {author} {\bibfnamefont {E.~V.}\ \bibnamefont {Zikiy}}, \bibinfo {author} {\bibfnamefont {I.~A.}\ \bibnamefont {Rodionov}},\ and\ \bibinfo {author} {\bibfnamefont {I.~S.}\ \bibnamefont {Besedin}},\ }\href@noop {} {\bibfield  {journal} {\bibinfo  {journal} {npj Quantum Inf.}\ }\textbf {\bibinfo {volume} {8}},\ \bibinfo {pages} {130} (\bibinfo {year} {2022})}\BibitemShut {NoStop}%
\bibitem [{\citenamefont {Walter}\ \emph {et~al.}(2017)\citenamefont {Walter}, \citenamefont {Kurpiers}, \citenamefont {Gasparinetti}, \citenamefont {Magnard}, \citenamefont {Poto\ifmmode~\check{c}\else \v{c}\fi{}nik}, \citenamefont {Salath\'e}, \citenamefont {Pechal}, \citenamefont {Mondal}, \citenamefont {Oppliger}, \citenamefont {Eichler},\ and\ \citenamefont {Wallraff}}]{Walter2017}%
  \BibitemOpen
  \bibfield  {author} {\bibinfo {author} {\bibfnamefont {T.}~\bibnamefont {Walter}}, \bibinfo {author} {\bibfnamefont {P.}~\bibnamefont {Kurpiers}}, \bibinfo {author} {\bibfnamefont {S.}~\bibnamefont {Gasparinetti}}, \bibinfo {author} {\bibfnamefont {P.}~\bibnamefont {Magnard}}, \bibinfo {author} {\bibfnamefont {A.}~\bibnamefont {Poto\ifmmode~\check{c}\else \v{c}\fi{}nik}}, \bibinfo {author} {\bibfnamefont {Y.}~\bibnamefont {Salath\'e}}, \bibinfo {author} {\bibfnamefont {M.}~\bibnamefont {Pechal}}, \bibinfo {author} {\bibfnamefont {M.}~\bibnamefont {Mondal}}, \bibinfo {author} {\bibfnamefont {M.}~\bibnamefont {Oppliger}}, \bibinfo {author} {\bibfnamefont {C.}~\bibnamefont {Eichler}},\ and\ \bibinfo {author} {\bibfnamefont {A.}~\bibnamefont {Wallraff}},\ }\href {https://doi.org/10.1103/PhysRevApplied.7.054020} {\bibfield  {journal} {\bibinfo  {journal} {Phys. Rev. Appl.}\ }\textbf {\bibinfo {volume} {7}},\ \bibinfo {pages} {054020} (\bibinfo {year} {2017})}\BibitemShut {NoStop}%
\bibitem [{\citenamefont {Swiadek}\ \emph {et~al.}(2024)\citenamefont {Swiadek}, \citenamefont {Shillito}, \citenamefont {Magnard}, \citenamefont {Remm}, \citenamefont {Hellings}, \citenamefont {Lacroix}, \citenamefont {Ficheux}, \citenamefont {Zanuz}, \citenamefont {Norris}, \citenamefont {Blais}, \citenamefont {Krinner},\ and\ \citenamefont {Wallraff}}]{Swiadek2024}%
  \BibitemOpen
  \bibfield  {author} {\bibinfo {author} {\bibfnamefont {F.}~\bibnamefont {Swiadek}}, \bibinfo {author} {\bibfnamefont {R.}~\bibnamefont {Shillito}}, \bibinfo {author} {\bibfnamefont {P.}~\bibnamefont {Magnard}}, \bibinfo {author} {\bibfnamefont {A.}~\bibnamefont {Remm}}, \bibinfo {author} {\bibfnamefont {C.}~\bibnamefont {Hellings}}, \bibinfo {author} {\bibfnamefont {N.}~\bibnamefont {Lacroix}}, \bibinfo {author} {\bibfnamefont {Q.}~\bibnamefont {Ficheux}}, \bibinfo {author} {\bibfnamefont {D.~C.}\ \bibnamefont {Zanuz}}, \bibinfo {author} {\bibfnamefont {G.~J.}\ \bibnamefont {Norris}}, \bibinfo {author} {\bibfnamefont {A.}~\bibnamefont {Blais}}, \bibinfo {author} {\bibfnamefont {S.}~\bibnamefont {Krinner}},\ and\ \bibinfo {author} {\bibfnamefont {A.}~\bibnamefont {Wallraff}},\ }\href {https://doi.org/10.1103/PRXQuantum.5.040326} {\bibfield  {journal} {\bibinfo  {journal} {PRX Quantum}\ }\textbf {\bibinfo {volume} {5}},\ \bibinfo {pages} {040326} (\bibinfo {year} {2024})}\BibitemShut {NoStop}%
\bibitem [{\citenamefont {Sunada}\ \emph {et~al.}(2024)\citenamefont {Sunada}, \citenamefont {Yuki}, \citenamefont {Wang}, \citenamefont {Miyamura}, \citenamefont {Ilves}, \citenamefont {Matsuura}, \citenamefont {Spring}, \citenamefont {Tamate}, \citenamefont {Kono},\ and\ \citenamefont {Nakamura}}]{Sunada2024}%
  \BibitemOpen
  \bibfield  {author} {\bibinfo {author} {\bibfnamefont {Y.}~\bibnamefont {Sunada}}, \bibinfo {author} {\bibfnamefont {K.}~\bibnamefont {Yuki}}, \bibinfo {author} {\bibfnamefont {Z.}~\bibnamefont {Wang}}, \bibinfo {author} {\bibfnamefont {T.}~\bibnamefont {Miyamura}}, \bibinfo {author} {\bibfnamefont {J.}~\bibnamefont {Ilves}}, \bibinfo {author} {\bibfnamefont {K.}~\bibnamefont {Matsuura}}, \bibinfo {author} {\bibfnamefont {P.~A.}\ \bibnamefont {Spring}}, \bibinfo {author} {\bibfnamefont {S.}~\bibnamefont {Tamate}}, \bibinfo {author} {\bibfnamefont {S.}~\bibnamefont {Kono}},\ and\ \bibinfo {author} {\bibfnamefont {Y.}~\bibnamefont {Nakamura}},\ }\href {https://doi.org/10.1103/PRXQuantum.5.010307} {\bibfield  {journal} {\bibinfo  {journal} {PRX Quantum}\ }\textbf {\bibinfo {volume} {5}},\ \bibinfo {pages} {010307} (\bibinfo {year} {2024})}\BibitemShut {NoStop}%
\bibitem [{\citenamefont {Spring}\ \emph {et~al.}(2024)\citenamefont {Spring}, \citenamefont {Milanovic}, \citenamefont {Sunada}, \citenamefont {Wang}, \citenamefont {van Loo}, \citenamefont {Tamate},\ and\ \citenamefont {Nakamura}}]{Spring2024}%
  \BibitemOpen
  \bibfield  {author} {\bibinfo {author} {\bibfnamefont {P.~A.}\ \bibnamefont {Spring}}, \bibinfo {author} {\bibfnamefont {L.}~\bibnamefont {Milanovic}}, \bibinfo {author} {\bibfnamefont {Y.}~\bibnamefont {Sunada}}, \bibinfo {author} {\bibfnamefont {S.}~\bibnamefont {Wang}}, \bibinfo {author} {\bibfnamefont {A.~F.}\ \bibnamefont {van Loo}}, \bibinfo {author} {\bibfnamefont {S.}~\bibnamefont {Tamate}},\ and\ \bibinfo {author} {\bibfnamefont {Y.}~\bibnamefont {Nakamura}},\ }\href@noop {} {\bibinfo {title} {Fast multiplexed superconducting qubit readout with intrinsic {Purcell} filtering}} (\bibinfo {year} {2024}),\ \Eprint {https://arxiv.org/abs/2409.04967} {arXiv:2409.04967 [quant-ph]} \BibitemShut {NoStop}%
\bibitem [{\citenamefont {Chatterjee}\ \emph {et~al.}(2024)\citenamefont {Chatterjee}, \citenamefont {Schwinger},\ and\ \citenamefont {Gao}}]{Chatterjeee2024}%
  \BibitemOpen
  \bibfield  {author} {\bibinfo {author} {\bibfnamefont {A.}~\bibnamefont {Chatterjee}}, \bibinfo {author} {\bibfnamefont {J.}~\bibnamefont {Schwinger}},\ and\ \bibinfo {author} {\bibfnamefont {Y.~Y.}\ \bibnamefont {Gao}},\ }\href@noop {} {\bibinfo {title} {Demonstration of enhanced qubit readout via reinforcement learning}} (\bibinfo {year} {2024}),\ \Eprint {https://arxiv.org/abs/2412.04053} {arXiv:2412.04053 [quant-ph]} \BibitemShut {NoStop}%
\bibitem [{\citenamefont {Catelani}\ \emph {et~al.}(2011)\citenamefont {Catelani}, \citenamefont {Koch}, \citenamefont {Frunzio}, \citenamefont {Schoelkopf}, \citenamefont {Devoret},\ and\ \citenamefont {Glazman}}]{Koch2011}%
  \BibitemOpen
  \bibfield  {author} {\bibinfo {author} {\bibfnamefont {G.}~\bibnamefont {Catelani}}, \bibinfo {author} {\bibfnamefont {J.}~\bibnamefont {Koch}}, \bibinfo {author} {\bibfnamefont {L.}~\bibnamefont {Frunzio}}, \bibinfo {author} {\bibfnamefont {R.~J.}\ \bibnamefont {Schoelkopf}}, \bibinfo {author} {\bibfnamefont {M.~H.}\ \bibnamefont {Devoret}},\ and\ \bibinfo {author} {\bibfnamefont {L.~I.}\ \bibnamefont {Glazman}},\ }\href {https://doi.org/10.1103/PhysRevLett.106.077002} {\bibfield  {journal} {\bibinfo  {journal} {Phys. Rev. Lett.}\ }\textbf {\bibinfo {volume} {106}},\ \bibinfo {pages} {077002} (\bibinfo {year} {2011})}\BibitemShut {NoStop}%
\end{thebibliography}%

\end{document}